\documentclass[twocolumn]{aastex7}
\usepackage{natbib}[1999/02/25]
\usepackage{booktabs}
\usepackage{url}
\usepackage{amsmath}
\usepackage{soul}
\usepackage{longtable}
\usepackage{MnSymbol}
\usepackage{CJKutf8}
\usepackage{psfrag,color}
\usepackage{multirow}
\usepackage{subfigure}
\usepackage{cancel}
\newcount\ppnum
\newcommand\ppnumber[1]{%
        \ppnum=#1\relax
        \ifnum\ppnum<0
                $-$%
                \ppnum=-\ppnum
        \fi
        \let\pptemp\empty
        \loop\ifnum\ppnum>999
                \count255=\ppnum
                \divide\ppnum by1000
                \count255=\numexpr \count255 - 1000*\ppnum \relax
                \edef\pptemp{,\ifnum\count255<100 0\ifnum\count255<10 0\fi\fi
                             \the\count255 \pptemp}%
        \repeat
        \the\ppnum
        \pptemp
}
\shorttitle{Bolometric Correction and Zero-Point Constants of Visual Magnitudes}
\shortauthors{Y\"{u}cel et al.}
\graphicspath{{./}{Graphs/}}
\begin{document}

\title{Empirical Bolometric Correction and Zero-Point Constants of Visual Magnitudes from High-Resolution Spectra}

\correspondingauthor{G\"{o}khan Y\"{u}cel}

\author[0000-0002-9846-3788]{G\"{o}khan Y\"{u}cel}
\altaffiliation{T\"{U}B{\.{I}}TAK-2218 Fellow}
\affiliation{Istanbul University, Faculty of Sciences, Department of Astronomy and Space Sciences, 34119, Istanbul, T\"{u}rkiye}
\email[show]{gokhannyucel@gmail.com}

\author[0000-0003-3510-1509, sname='Bilir']{Sel\c{c}uk Bilir}
\affiliation{Istanbul University, Faculty of Sciences, Department of Astronomy and Space Sciences, 34119, Istanbul, T\"{u}rkiye}
\email{sbilir@istanbul.edu.tr}

\author[0000-0002-3125-9010, sname='Bak{\i}\c{s}']{Volkan Bak{\i}\c{s}}
\affiliation{Akdeniz University, Faculty of Sciences, Department of Space Sciences and Technologies, 07058, Antalya, T\"{u}rkiye}
\email{vbakis@akdeniz.edu.tr}

\author[0000-0003-1883-6255, sname='Eker']{Zeki Eker}
\affiliation{Akdeniz University, Faculty of Sciences, Department of Space Sciences and Technologies, 07058, Antalya, T\"{u}rkiye}
\email{eker@akdeniz.edu.tr}

	
	

\begin{abstract}
\noindent 
A method of obtaining bolometric corrections ($BC_{\rm V}$) from observed high-resolution, high- \textit{S/N} spectra is described. The method is applied to spectra of 128 stars collected from the literature with well-determined effective temperatures ($T_{\rm eff}$) with $S_\lambda(V)$ transparency profiles of Bessell and Landolt. Computed $BC_{\rm V}$ are found accurate within several milimagnitudes and the effect of different $S_\lambda(V)$ is found to be no more than 0.015 mag. Measured visual to bolometric ratio ($L_{\rm V}/L$) from the sample spectra and classically determined $BC_{\rm V}$ from bolometric ($M_{\rm Bol}$) and visual ($M_{\rm V}$) absolute magnitudes helped us to determine the zero-point constant ($C_{\rm 2}$) of the $BC_{\rm V}$ scale. Determined $C_{\rm 2}$ for each star for each $S_\lambda(V)$ profile revealed $C_{\rm 2} = 2.3653\pm0.0067$ mag if $S_\lambda(V)$ profile of Bessell is used, and $C_{\rm 2} = 2.3826\pm0.0076$ mag if $S_\lambda(V)$ profile of Landolt is used. Expanding $C_{\rm Bol} = 71.197425 ...$ mag and $c_{\rm Bol} = -18.997351 ...$ mag announced by IAU2015GARB2, and using definition of $C_{\rm 2} = C_{\rm Bol}-C_{\rm V} = c_{\rm Bol}-c_{\rm V}$, where capital $C$ is for the absolute and small $c$ is for the apparent, subscripts indicating bolometric and visual, the zero-point constants: $C_{\rm V} = 68.8321\pm0.0067$ mag and $c_{\rm V} = -21.3627\pm0.0067$ mag, if $L_{\rm V}$ and are in SI units, were determined corresponding to $S_\lambda(V)$ of Bessell. The zero-point constants corresponding to $S_\lambda(V)$ of Landolt are smaller, but the difference is not more than 0.02 mag. Typical and limiting accuracies for predicting a stellar luminosity from an apparent magnitude and a distance are analyzed.

\end{abstract}
	
	

%----------------------------------------------------------------------------------------------
 
\section{Introduction} \label{sec:intro}
Although the concept of ``bolometric correction" (\textit{BC})  was introduced as an instrument to be used to establish the stellar effective temperature ($T_{\rm eff}$) scale, standard stellar bolometric corrections today are on the way to take the role of providing the most accurate stellar luminosities (percent level) \citep{Eker2025}. A century-long quest started when \cite{Kuiper1938} defined the \textit{BC} of a star as:

\begin{equation}
    BC = M_{\rm Bol}-M_{\rm V} = m_{\rm Bol}-V,
    \label{equ:1}
\end{equation}

\noindent where \textit{BC} appears simply as the difference between the star’s bolometric and visual magnitudes. Because this equation could also be written as $M_{\rm Bol}=M_{\rm V}+BC$, and $m_{\rm Bol}=V+BC$, where \textit{BC} appears as a term if it is added to the visual, one obtains the bolometric, that is, the total brightness of the star at all wavelengths from zero to infinity in both absolute or apparent regimes, \cite{Kuiper1938} named it “bolometric correction”. This name was well-established and has been used throughout the century, and still remains in use today. However, implying that there must not be a zero-point constant for the \textit{BC} scale \citep{Torres2010} or if there is one, it must be less than zero; the verbal definition of \textit{BC} according to Equation~(\ref{equ:1}) is paradoxical or ill posed to indicate: $L_{\rm V}=L\times10^{(BC/{2.5})}$, since if $BC>0$, $L_{\rm V}$ is nonphysical, where $L$ and $L_{\rm V}$ are the total and partial (visual) luminosities of the star, from which perplexing paradigms 1) “the bolometric magnitude of a star ought to be brighter than its visual magnitude”, 2)“bolometric corrections must always be negative” and 3) “the zero point of bolometric corrections are arbitrary” were emerged \citep{Eker2025}. 

Because of the ill-posed nature of the original definition of Kuiper, numerous \textit{BC} tables \citep{Torres2010, Eker2021} and $BC-T_ {\rm eff}$ relations \citep{Flower1996, Eker2020} were produced to compete. First, it was because almost half of the published tables \citep{Kuiper1938, McDonald1952, Popper1959, Wildey1963, Hayes1978, Habets1981, Cox2000, Pecaut2013} produced by the authors obey the three paradigms, thus all \textit{BC} values are negative while the others  \citep{Johnson1964, Johnson1966, Code1976, Flower1977, Flower1996, Bessell1998, Sung2013, Casagrande2018, Eker2020} allow a limited number of positive \textit{BC}, that is, disobeying the paradigms.  

Another serious problem was that, because of the arbitrariness attributed to the zero-point constant of \textit{BC} scales (paradigm 3), a star most likely could be found with more than one or several \textit{BC} assigned to it. Numerous competing \textit{BC} for a star, however, require multiple competing bolometric absolute magnitudes for the same star as indicated by Equation~(\ref{equ:1}), despite it having a single absolute visual magnitude. The numerous inconsistent absolute bolometric magnitudes, then, imply numerous inconsistent stellar luminosities ($L$) for the same star due to the following relation.

\begin{equation}
    M_\mathrm{Bol} = M_\mathrm{Bol,\odot}-2.5 \times \log{L/L_\odot}
    \label{equ:2}
\end{equation}
This equation also introduces additional uncertainty because different users tend to use dissimilar values of $M_{\rm Bol,\odot}$ and $L_\odot$.

The error contribution of a non-standard \textit{BC} on a predicted $L$ was estimated to be $10\%$ or more according to \cite{Torres2010}. When \cite{Andersen1991} and \cite{Torres2010b} collected the most accurate masses ($M$) and radii ($R$) of the Detached Double-lined Eclipsing Binaries (DDEB), which are accurate within $3\%$, and \cite{Masana2006} were estimating the errors of the effective temperatures $1\%-2\%$,  a $10\%$ or more uncertainty solely from a non-standard \textit{BC} were annoying. Error contributions of apparent magnitudes and trigonometric parallaxes were reduced greatly to be about $5\%$ or less after Hipparcos mission operated in 1989-1993 \citep{ESA1997} for nearby stars up to 8-9 magnitudes, and approximately few percent or faint for the stars up to 21st mag during {\it Gaia} mission \citep{Gaia_DR3} operated in 2013-2025. That is, the error contribution of a non-standard \textit{BC} reached an intolerable level when the IAU issued a resolution \citep[hereafter IAU2015GARB2,][]{Mamajek2015} in the XXIX’th International Astronomical Union General Assembly in Honolulu in 2015.

The IAU 2015 General Assembly was also aware of the problems associated with Equation (\ref{equ:2}); thus, 

\begin{equation}
    M_\mathrm{Bol} = -2.5 \times \log{L} + C_\mathrm{Bol}
    \label{equ:3}
\end{equation}

\noindent was announced to replace it, where the zero-point constant of absolute bolometric magnitudes: $C_{\rm Bol} = 71.197425$…, if $L$ is in SI units, was fixed to resolve the problems associated with nonstandard tabulations of \textit{BC} and to avoid the variable Sun even though the solar variability is too small ($\sim1 \%$) to be felt within a long period ($\sim11$ years) known as the solar cycle \citep{Kopp2014}.  

The revolutionary status of the resolution \citep{Eker2022} stayed unnoticed for about seven years. Then, it was used as an argument against the arbitrariness of the \textit{BC} scale by \cite{Eker2021}, who defined the standard \textit{BC} of a star as the difference between its $M_{\rm Bol}$ and $M_{\rm V}$ if $M_{\rm Bol}$ was calculated by Equation~(\ref{equ:3}) using the Stefan-Boltzmann law, $L= 4\pi R^2 \sigma T^4$, and if $M_{\rm V}$ came from the most accurate parallax and apparent visual brightness corrected for interstellar extinction. Soon after, \cite{Eker2021b} defined the standard luminosity as the $L$ value calculated through Equation~(\ref{equ:3}) using a $M_{\rm Bol}$, value estimated as $M_{\rm Bol} = M_{\rm V} + BC$, with a standard \textit{BC}. 

When reviewing the three methods of estimating $L$ of a star in the era after \textit{Gaia}, the direct method using $R$ and $T_{\rm eff}$, was found the most accurate with limiting and typical accuracies of $2.5 \%$ and $8.2\% - 12.2 \%$, respectively, by \cite{Eker2021b}. The other two indirect methods, one requiring a pre-determined standard \textit{BC} and the other requiring an MLR (mass-luminosity relation), were found to provide less accurate $L$ with a typical accuracy of $13.7\%-20.2\%$ or less \citep{Eker2024}. 

\cite{Bakis2022} developed a method to improve the accuracy of the standard $L$ of a star. Independently determined numerous multi-band apparent magnitudes, if properly corrected for interstellar extinction would provide numerous $M_{\rm Bol}$ values with a single reliable trigonometric parallax if multi-band standard $\textit{BC}_\xi$ values were available to calculate numerous $M_{\rm Bol} (\xi)=M_\xi+BC_\xi$, where $\xi$ is one of the photometric bands. Independently determined $M_{\rm Bol}$ values were then combined for a mean value, which was plugged into Equation~(\ref{equ:3}) to have a more accurate $L$. The standard error of the mean was propagated to be the uncertainty of the $L$.

\cite{Eker2023} tested the method and the standard multiband $BC_\xi$ values for the main-sequence stars by recovering $L$ and $R$ of the most accurate 341 single host stars (281 dwarfs, 40 subgiants, 19 giants, and one pre-main-sequence star). It is the first time in the history of astrophysics that there is a method to calculate an empirical $L$ value for a star, which is much more accurate than the direct method can provide. This is, of course, one of the outstanding results of the resolution (IAU2015GARB2) issued by IAU in 2015 and the definition of the standard \textit{BC} by \cite{Eker2021}. 

In addition to these improvements that motivated us for this study, we were further stimulated by \cite{Eker2021b}, who claimed additional enhancements, up to 1\%, and possibly more if the unique \textit{BC} of a star is measured directly from its observed spectrum. Therefore, the primary intention of this study is to describe a method of obtaining an empirical \textit{BC} of a star from its spectrum and zero-point constants of visual (absolute/apparent) magnitudes from 128 high resolution ($R > 25\,000$) and high signal-to-noise ($S/N >102$) spectra, which we collected from the literature.   

\section{Data} \label{sec:Data}

\begin{table*}[ht]
    \caption{Information about spectrographs and spectral libraries.}
    \centering
    \begin{tabular}{llcccrc}
    \hline
        Order & Instrument & Resolving Power & Wavelength Range (\AA) & \textit{S/N} & $N$ & Source \\ 
        \hline
        1 & MELCHIORS& 85\,000           & 3\,800-9\,000  & (128-349]        & 66 & 1\\ 
        2 & PEPSI    & 200\,000-270\,000 & 3\,830-9\,120  & (184-2\,947]     & 36 & 2\\ 
        3 & FIES     & 25\,000-46\,000   & 3\,700-7\,300  & (102-267]        &  8 & 3\\
        4 & HERMES   & 85\,000           & 3\,800-9\,000  & (112-208]        &  6 & 3\\ 
        5 & FEROS    & 48\,000           & 3\,526-9\,215  & (177-315]        &  5 & 3\\ 
        6 & ESPaDOnS & 68\,000           & 3\,700-10\,500 & (228-583]        &  3 & 4, 5, 6\\ 
        7 & NARVAL   & 65\,000           & 3\,700-10\,000 & (1\,054-1\,483]  &  2 & 7  \\
        8 & FTS      & 348\,000-522\,000 & 2\,960-13\,000 & (2\,000-3\,000]  &  1 & 8 \\ 
        \hline
    \end{tabular}\\
     (1) \citet{Melchiors2024}, (2) \citet{Strassmeier2018}, (3) \citet{Simon2020}, (4) \cite{Espadons}, \\
    (5) \cite{Donati1997}, (6) \cite{Petit2014}, (7) \cite{Auriere2003}, (8) \citet{Kurucz1984} \\
        \label{tab:1}
\end{table*}

Various spectrum libraries were visited to build a star list to be used for computing BC and zero-point constants of visual magnitudes from high-resolution spectra. The selection criteria were simple; if a single star has a spectrum without noticeable emission feature within the wavelength range at least to cover the range of the $V$ filter with high $S/N$, typically $>100$, and high resolution, typically $R>20\,000$, and well established $T_{\rm eff}$ in literature, it is included in the list. We have been careful in collecting stars with a wide range of $T_{\rm eff}$ values belonging to different luminosity classes as much as possible.   

Information about libraries and spectrographs is given in Table~\ref{tab:1}. Instruments, some critical information of the spectra, the number of spectra chosen, and a reference to give further details are indicated in the table. 

\subsection{Instruments and Spectral Libraries}

\subsubsection{HERMES}
As can be seen in Table~\ref{tab:1}, most spectra in our list were obtained with the HERMES (High-Efficiency and high-Resolution Mercator Echelle Spectrograph) spectrograph \citep{Raskin2011} attached to the 1.2m Mercator telescope at the Observatorio del Roque de Los Muchachos in La Palma. It has the capability of obtaining a spectrum with a resolving power up to 85\,000, covering wavelengths between 3\,800 \AA~ and 9\,000 \AA. There are two spectral libraries that we have collected spectra from, MELCHIORS\footnote{\url{https://royer.se/melchiors.html}} (Mercator Library of High Resolution Stellar Spectroscopy) \citep{Royer2014} and the IACOB spectroscopic database\footnote{\url{https://research.iac.es/proyecto/iacob/iacobcat/}} \citep{Simon2020}. 

\subsubsection{PEPSI}
The second-largest number of spectra in our sample is from the Potsdam Echelle Polarimetric and Spectroscopic Instrument \citep{Strassmeier2015} that is attached to the Large Binocular Telescope (LBT) at the Mount Graham International Observatory in Arizona. It has the capability of obtaining spectra with resolving power up to 270\,000, covering wavelengths between 3\,830 and 9\,120 \AA. We have collected spectra from \cite{Strassmeier2018}\footnote{\url{https://pepsi.aip.de/?page_id=552}}.

\subsubsection{FIES}

We have collected eight spectra from the high-resolution FIbre-fed Echelle Spectrograph (FIES) \citep{Telting2014} that is attached to 2.59m The Nordic Optical Telescope \citep{Djupvik2010} at the Observatorio del Roque de Los Muchachos in La Palma. It has a resolving power up to 67\,000, covering wavelengths between 3\,700 and 7\,300 \AA. We have collected spectra from the IACOB spectroscopic database.

\subsubsection{FEROS}

Five spectra of our sample were taken via Fiberfed Extended Range Optical Spectrograph (FEROS) \citep{Kaufer1999} that is attached to the 2.2m MPG/ESO telescope at the La Silla Observatory. The resolving power is up to 48\,000 with a wavelength coverage between 3\,500 and 9\,200 \AA. We made use of the spectra available in the IACOB spectroscopic database.

\subsubsection{ESPaDOnS}

The three spectra in our study are from the Echelle spectropolarimetric device for the observation of stars (ESPaDOnS) \citep{Manset2003} that is attached to the 3.6m Canada-France-Hawaii Telescope in Maunakea, Hawaii. It has a capability of resolving power of up to 68\,000 with a wavelength coverage between 3\,700 and 10\,500 \AA. The spectra provided by \citet{Espadons} and Polarbase Observatory \citep{Donati1997, Fossati2011}.

\subsubsection{NARVAL}

Among our minimal sample of spectra, two spectra were obtained using the NARVAL spectrograph \citep{Auriere2003}, which is mounted on the 2.03m telescope at the Pic du Midi Observatory. It has a wavelength coverage from 3\,700 to 10\,000 {\AA} with a resolving power of 65\,000.

\subsubsection{FTS}

The Sun, considered as a star, is also included in our list, represented by a solar spectrum obtained with the Fourier Transform Spectrometer (FTS) attached to the McMath Solar Telescope at Kitt Peak National Observatory. Resolution of the spectrum changes between 348\,000 and 522\,000, in the ultraviolet region and the infrared region, respectively. Total wavelength coverage of the spectrum is between 2\,960 and 13\,000 \AA~\citep{Kurucz1984}.

\subsection{Photometric Data for Absolute $M_{\rm Bol}$ and $M_{\rm V}$}

Finding a reliable standard $BC$ with the classical method, Equation~(\ref{equ:1}), for single stars requires accurate $M_{\rm Bol}$ values that should be obtained by using Equation~(\ref{equ:3}). First, the most reliable value of $L$ for a star could be determined if sufficiently accurate $T_{\rm eff}$ and $R$ are available. Therefore, while building our star list, we have selected stars with $T_{\rm eff}$ determined spectroscopically by a method called model atmosphere fitting to the observed stellar spectra. It is possible to estimate the radius ($R$) of a star from its model atmosphere parameter surface gravity ($\log g$) if its mass ($M$) is available. Unfortunately, reliable stellar masses are possible only for binaries or multiple stars via Kepler's third law. Kepler's third law cannot apply to single stars. Therefore, we have preferred to estimate $R$ and its uncertainty via SED analysis. HD\,16440 is the only star whose $T_{\rm eff}$ and $R$ are determined by the SED analysis.

\subsection{SED Analysis for $R$ and \(A_{\rm V}\)} 

We have taken the SED modeling approach of \cite{Bakis2022} not only to estimate the radius ($R$) and its uncertainty required for the luminosity and its error for a star, and then for its absolute bolometric magnitude ($M_{\rm Bol}$) and its uncertainty through Equation~(\ref{equ:3}), but also for estimating interstellar extinctions ($A_{\rm V}$) together with its uncertainty which is needed for a reliable absolute visual magnitude ($M_{\rm V}$) and its uncertainty.  At last, the standard {\it BC} of the sample stars could be calculated according to Equation~(\ref{equ:1}).

%-----------------------------------------------------------------
% Figure 2
\begin{figure*}
    \centering
    \includegraphics[width=0.45\linewidth]{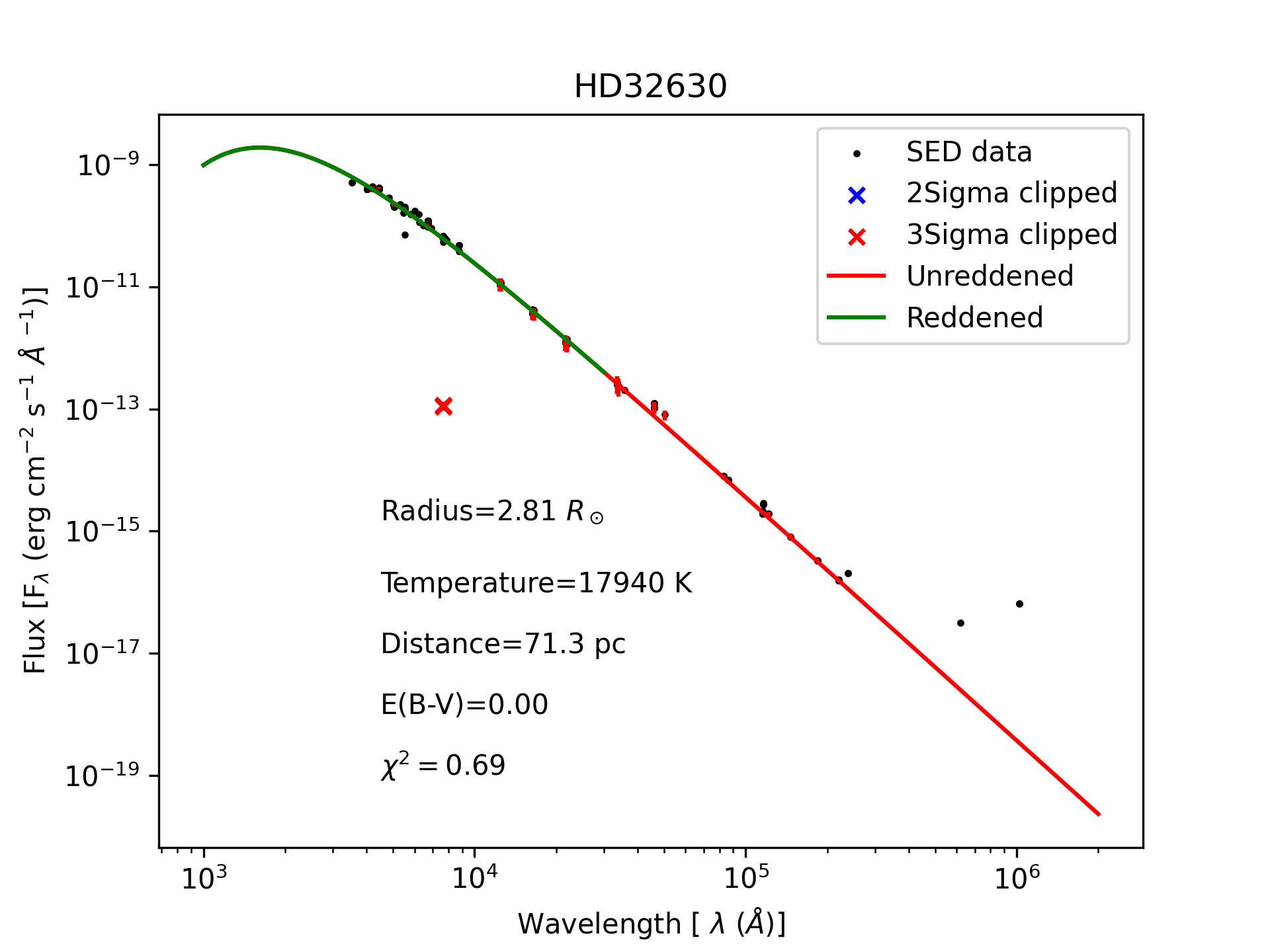}
    \includegraphics[width=0.45\linewidth]{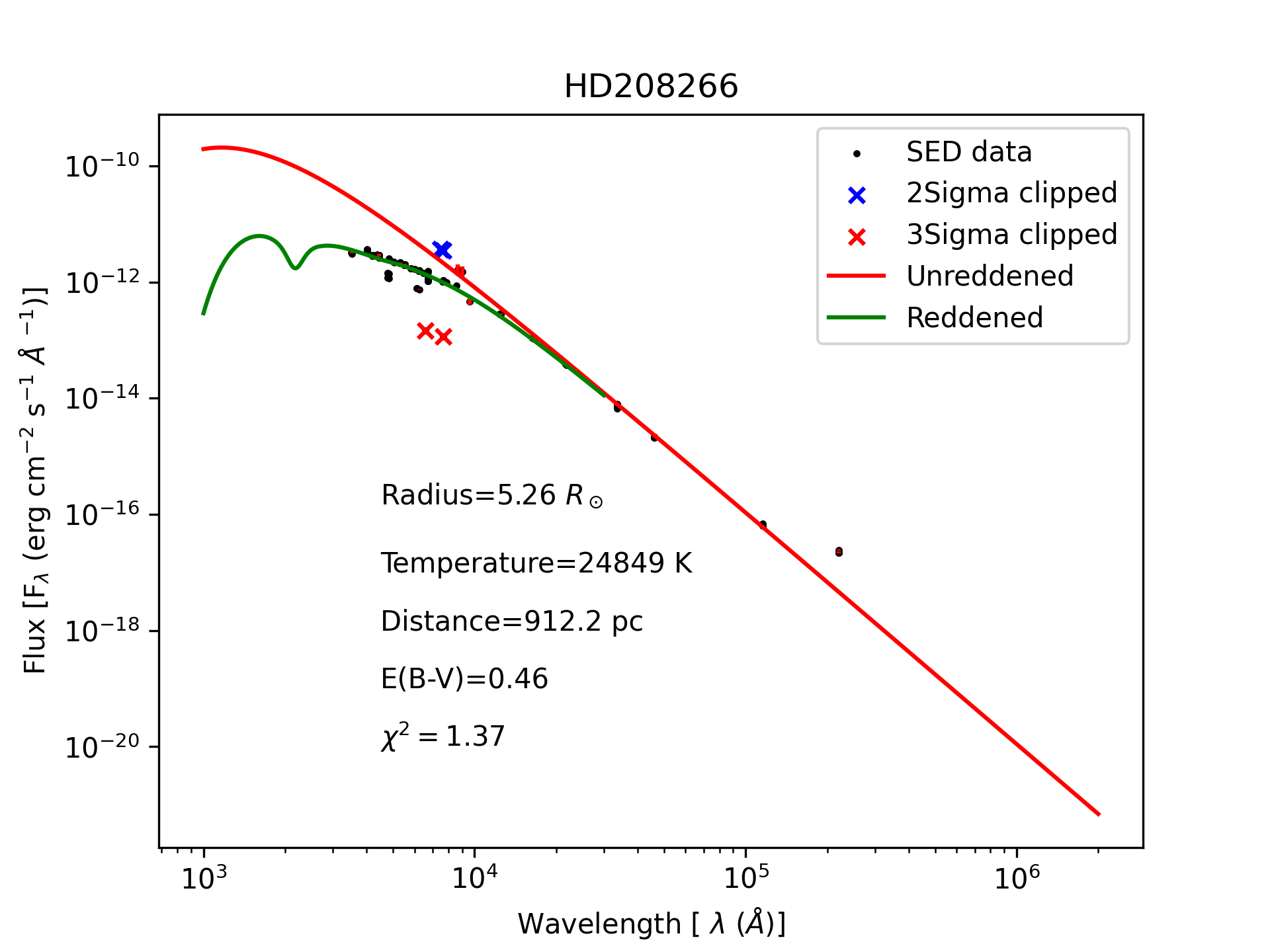}
    \includegraphics[width=0.45\linewidth]{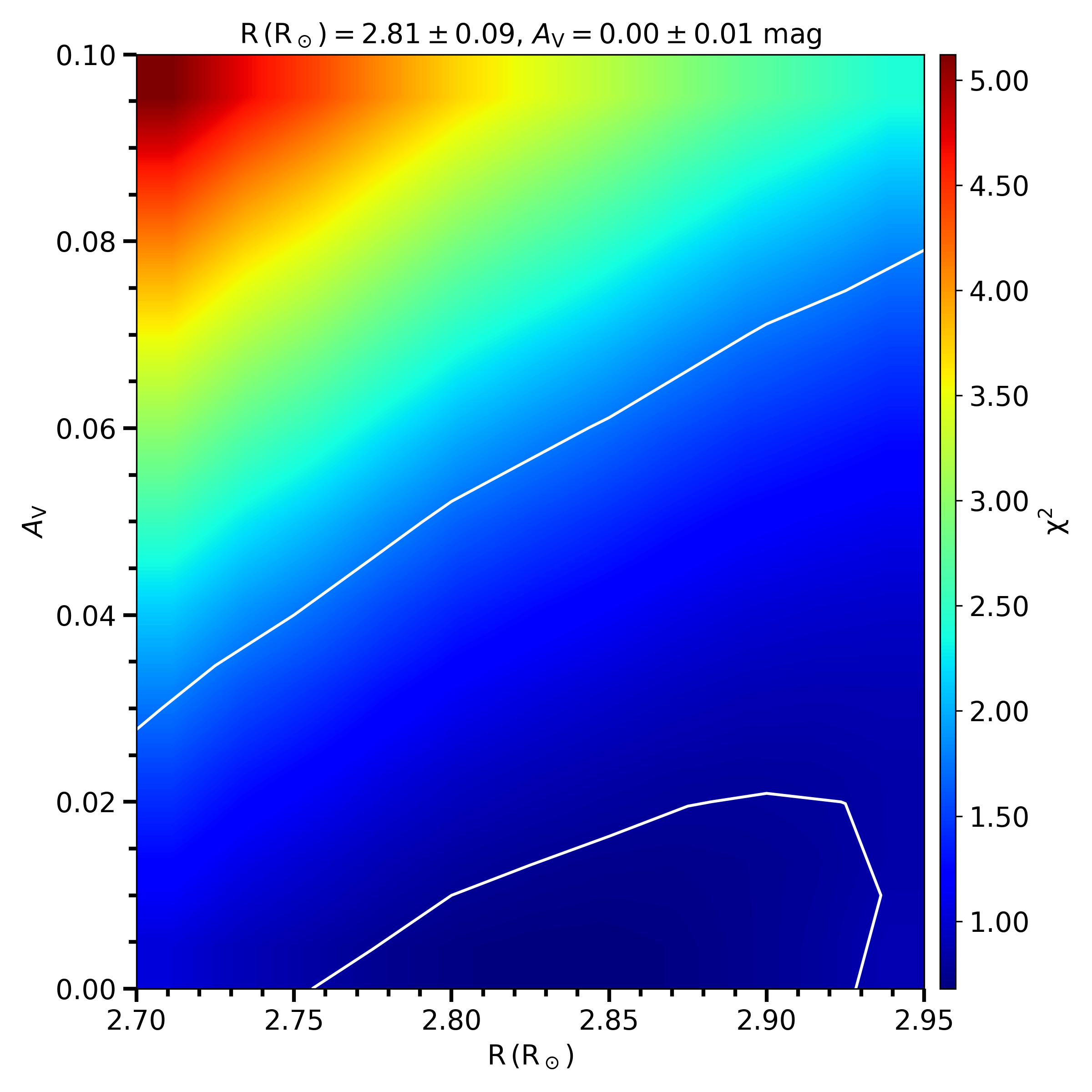}
    \includegraphics[width=0.45\linewidth]{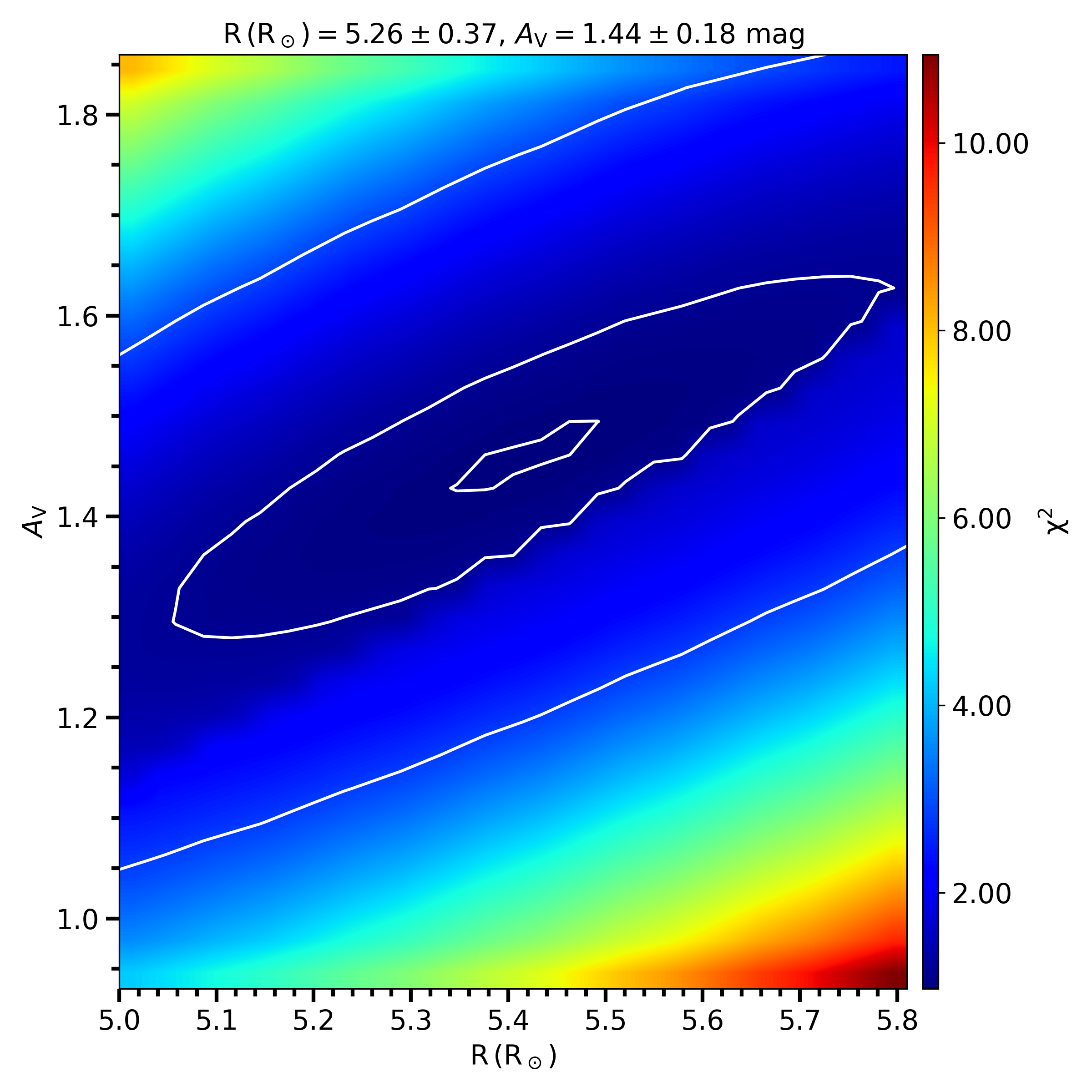}
    \caption{\textit{Top panel:} SED models of two stars (HD\,32630 and HD\,208266) with different interstellar extinctions. \textit{Bottom panel:} $\chi^2$ map of models given in the top panel.}
    \label{fig:sed}
\end{figure*}
%-----------------------------------------------------------------

The estimated $R$ are listed in Table \ref{tab:2} together with calculated $L$ in solar/SI units and corresponding $M_{\rm Bol}$ together with associated errors. Interstellar extinctions and errors, however, are listed in Table \ref{tab:3} among the other observational parameters and associated uncertainties, which were involved in computing  $M_{\rm V}$. The apparent visual magnitudes and uncertainties in columns 3 and 4, respectively, were taken from the SIMBAD database. The trigonometric parallaxes and errors in columns 5 and 6, respectively, are taken from {\it Gaia} DR3 \citep{Gaia_DR3} unless not available; then {\it Hipparcos} trigonometric parallaxes \citep{ESA1997} and errors were preferred. The source of parallaxes and errors is indicated in column 7. Absolute visual magnitudes ($M_{\rm V}$) and errors corrected for interstellar extinction are given in the two rightmost columns, respectively.

%-----------------------------------------------------------------

The determination of a radius ($R$) and an interstellar extinction ($A_{\rm V}$) for a star is demonstrated in Figure \ref{fig:sed} by the SED analysis of the two stars HD\,208266 and HD\,32630 with and without interstellar extinction, respectively. It is clear in Figure~\ref{fig:sed} on the SED of HD\,208266 that there are two solid curves, one representing the unreddened and the other representing the reddened SED that appears to fit observed spectrophotometric flux data from the SIMBAD database \citep{Wenger2000}. The unreddened SED is calculated by

\begin{equation}
    f_\lambda^0=\frac{R^2}{d^2}\pi B_\lambda(T_{\rm eff})
    \label{equ:blackbody}
\end{equation}
where $R$ is the radius, $d$ is the distance of the star, thus \(R^2/d^2\) is the dilution factor for the surface flux \(\pi B_\lambda(T_{\rm eff})\) of the star per unit wavelength. Consequently, $f_\lambda^0$ is the flux that is reaching the telescope if there is no atmospheric and interstellar extinction. This first-order approximation for a SED modeling approach by \cite{Bakis2022} assumes that the wavelength-dependent intensity \( B_\lambda(T_{\rm eff})\) represented by the Planck function is uniform over the solid angle \(\pi R^2/d^2\). The unreddened SED is reddened by adjusting \(E(B-V)\) of the system until a best-fitting reddened SED (\(f_{\lambda}\)) is obtained to fit the flux data, shown by the symbols on Figure \ref{fig:sed}, using the reddening model of \cite{Fitzpatrick1999}. \(R(\lambda)=A_\lambda/E(B-V)\) relations were used, among them \(R(V)=3.1\) adopted for $A_{\rm V}$ \citep{Cardelli1989}. As soon as the best-fitting reddened SED (\(f_{\lambda}\)) is obtained, the visual band extinctions (\(A_{\rm V}\)) of the sample stars were computed by taking the following integrals numerically according to the formula given by \cite{Bakis2022}. 

\begin{equation}
    A_{\rm V}=2.5\times\log\frac{\int_{0}^{\infty} S_\lambda (V)f_\lambda^0 \,d\lambda}{\int_{0}^{\infty} S_\lambda (V)f_0 \,d\lambda}
    \label{equ:reddening}
\end{equation}
where \(S_\lambda(V)\) is the transition profile of the visual filter. If $f_{\lambda}^0= f_\lambda$, it is clear that \(A_{\rm V}=0\) mag, which is the case with the star named HD\,32630 in Figure \ref{fig:sed} where the reddened and unreddened SEDs overlap. The spectrophotometric data from the SIMBAD database confine both $R$ and $A_{\rm V}$ values recorded in Tables~\ref{tab:2} and \ref{tab:3}, respectively.

\subsection{Spectroscopic data related to $L_{\rm V}/L$}

With a known transparency profile of a filter, let it be the visual filter expressed as $S_\lambda(V)$, it is possible to calculate the fractional luminosity (or visual to bolometric luminosity ratio) $L_{\rm V} / L$ of a star from its observed spectrum at least spanning the wavelength range of the visual filter. Actually, a fractional luminosity at a filter is the prime parameter for a star to its absolute and apparent magnitudes, as well as its $BC$ at various bands.

Being able to determine stellar $L_{\rm V}/L$ independently from photometric and spectroscopic data allowed us to determine the zero-point constant of the $BC_{\rm V}$ scale empirically by the help of Equation (\ref{equ:3}) containing the value of $C_{\rm Bol}$ from IAU2015GARB2 first and then to obtain empirical $BC_{\rm V}$ of 128 stars using individual spectroscopic $L_{\rm V}/L$ values and the newly determined zero-point constant for the $BC_{\rm V}$ scale.

%Tablo 2
\begin{longtable*}{llcclccccccc}

    \caption{Observational parameters for calculating absolute bolometric magnitudes ($M_{\rm Bol}$) of the sample stars.} \\
    \hline
    Order & Star & $T_{\rm eff}$ &  err & Reference & $R$ & err & $L$ & $L$ & err & $M_{\rm Bol}$ & err \\ 
 
         & & (K) & (\%) & & ($R_\odot$) & (\%) & (W) &  ($L_\odot$) & (\%) & \multicolumn{2}{c}{(mag)} \\  \hline
         \endfirsthead
    \multicolumn{12}{c}%
{{\bfseries \tablename\ \thetable{}} -- continued from previous page} \\ 
\hline  
    Order & Star & $T_{\rm eff}$ &  err & Reference & $R$ & err & $L$ & $L$ &  err & $M_{\rm Bol}$ & err\\ 
 
         & & (K) & (\%) & & ($R_\odot$) & (\%) & (W) & ($L_\odot$) & (\%) & \multicolumn{2}{c}{(mag)} \\ \hline
\endhead
\hline \multicolumn{12}{c}{{Continued on next page}} \\ \hline
\endfoot
\hline 
\endlastfoot
    \hline
        1 &     Sun     & 5772 & ~ &      \citet{Prsa2016}       & 1 & ~ & 3.83E+26 & 1 & ~ & 4.740 & ~ \\ 
        2 &     HD1279     & 13300 & 0.9 &      \citet{Monier2023}       & 5.73 & 5.0 & 3.54E+29 & 926 & 10.7 & -2.676 & 0.116 \\ 
        3 &     HD1404     & 8840 & 2.0 &      \citet{Hillen2012}       & 2.06 & 5.8 & 8.89E+27 & 23 & 14.0 & 1.325 & 0.152 \\ 
        4 &     HD1439     & 9640 & 1.3 &       \citet{Royer2014}       & 3.47 & 4.6 & 3.58E+28 & 94 & 10.5 & -0.188 & 0.114 \\ 
        5 &     HD2729     & 14125 & 6.8 &      \citet{SimonDiaz2017}       & 3.27 & 4.5 & 1.47E+29 & 384 & 28.7 & -1.720 & 0.311 \\ 
        ... & ... & ... & ... & ... & ... & ... & ... & ... & ... & ... & ... \\
        124 &     HD220009     & 4227 & 1.8 &      \citet{Soubiran2024}    & 23.9 & 6.3 & 6.29E+28 & 164 & 14.6 & -0.799 & 0.158 \\ 
        125 &     HD220825     & 10228 & 3.7 &      \citet{Prugniel2011}    & 1.66 & 3.0 & 1.04E+28 & 27 & 15.8 & 1.150 & 0.172 \\ 
        126 &     HD222173     & 11800 & 4.2 &      \citet{Bailey2013}    & 5.57 & 2.3 & 2.08E+29 & 542 & 17.6 & -2.096 & 0.191 \\ 
        127 &     HD222661     & 11108 & 3.4 &      \citet{David2015}    & 1.82 & 6.0 & 1.74E+28 & 45 & 18.1 & 0.599 & 0.197 \\ 
        128 &     HD222762     & 12828 & 0.8 &      \citet{Huang2010}    & 6.43 & 7.4 & 3.86E+29 & 1009 & 15.1 & -2.770 & 0.164\label{tab:2}
\end{longtable*}

%-----------------------------------------------------------------
%Tablo 3

\begin{longtable*}{llccrcccccc}

    \caption{Observational parameters for calculating absolute visual magnitudes $(M_{\rm V})$ of the sample stars.} \\
    \hline
    Order & Star & $V$ & err & \multicolumn{1}{c}{$\varpi$} & err & Source & $A_{\rm V}$ & err & $M_{\rm V}$ & err \\ 
 
         & & \multicolumn{2}{c}{(mag)} & \multicolumn{1}{c}{(mas)} & (\%) & & \multicolumn{2}{c}{(mag)} & \multicolumn{2}{c}{(mag)} \\ \hline
        \endfirsthead
    \multicolumn{11}{c}%
{{\bfseries \tablename\ \thetable{}} -- continued from previous page} \\ 
\hline  
    Order & Star & $V$ & err & \multicolumn{1}{c}{$\varpi$} & err & Source & $A_{\rm V}$ & err & $M_{\rm V}$ & err \\ 
 
         & & \multicolumn{2}{c}{(mag)} & \multicolumn{1}{c}{(mas)} & (\%) & & \multicolumn{2}{c}{(mag)} & \multicolumn{2}{c}{(mag)} \\ \hline
\endhead
\hline \multicolumn{11}{c}{{Continued on next page}} \\ \hline
\endfoot
\hline 
\endlastfoot
    \hline
        1 &  Sun  & -26.760 & 0.030 & ~ & ~ & ~ & 0.000 & ~ & 4.810 & 0.030 \\ 
        2 &  HD1279  & 5.764 & 0.014 & 2.8490 & 1.81 & $Gaia$ & 0.031 & 0.093 & -1.994 & 0.102 \\ 
        3 &  HD1404  & 4.520 & 0.010 & 23.2542 & 0.78 & $Gaia$ & 0.000 & 0.031 & 1.353 & 0.037 \\ 
        4 &  HD1439  & 5.875 & 0.009 & 6.6475 & 1.23 & $Gaia$ & 0.013 & 0.031 & -0.025 & 0.042 \\ 
        5 &  HD2729  & 6.165 & 0.010 & 3.8917 & 1.00 & $Gaia$ & 0.000 & 0.093 & -0.884 & 0.096 \\ 
        ... & ... & ... & ... & ... & ... & ... & ... & ... & ... & ... \\
        124 &  HD220009  & 5.069 & 0.009 & 9.0926 & 1.26 & $Gaia$ & 0.000 & 0.093 & -0.138 & 0.097 \\ 
        125 &  HD220825  & 4.940 & 0.010 & 20.3154 & 0.48 & $Gaia$ & 0.000 & 0.031 & 1.479 & 0.034 \\ 
        126 &  HD222173  & 4.290 & 0.010 & 6.4313 & 2.14 & $Gaia$ & 0.000 & 0.016 & -1.669 & 0.050 \\ 
        127 &  HD222661  & 4.484 & 0.009 & 20.8948 & 0.76 & $Gaia$ & 0.000 & 0.124 & 1.084 & 0.125 \\ 
        128 &  HD222762  & 6.630 & 0.010 & 1.9809 & 1.49 & $Gaia$ & 0.217 & 0.186 & -2.103 & 0.189
        \label{tab:3}
\end{longtable*}

%---------------------------------------------------------------

The methods for obtaining the zero-point constant for the $BC_{\rm V}$ scale, first, and then how to obtain individual spectroscopic $BC_{\rm V}$ from observed spectra are described in the appendix. Here we demonstrate and explain how to get solar $L_{\rm V}/L$ value from its high-resolution ($\Delta\lambda/\lambda\approx 350000-500000$), high (1349) \textit{S/N}  ratio solar spectrum from \citet{Kurucz1984} as an example.

\begin{figure*}
    \centering
    \includegraphics[width=\linewidth]{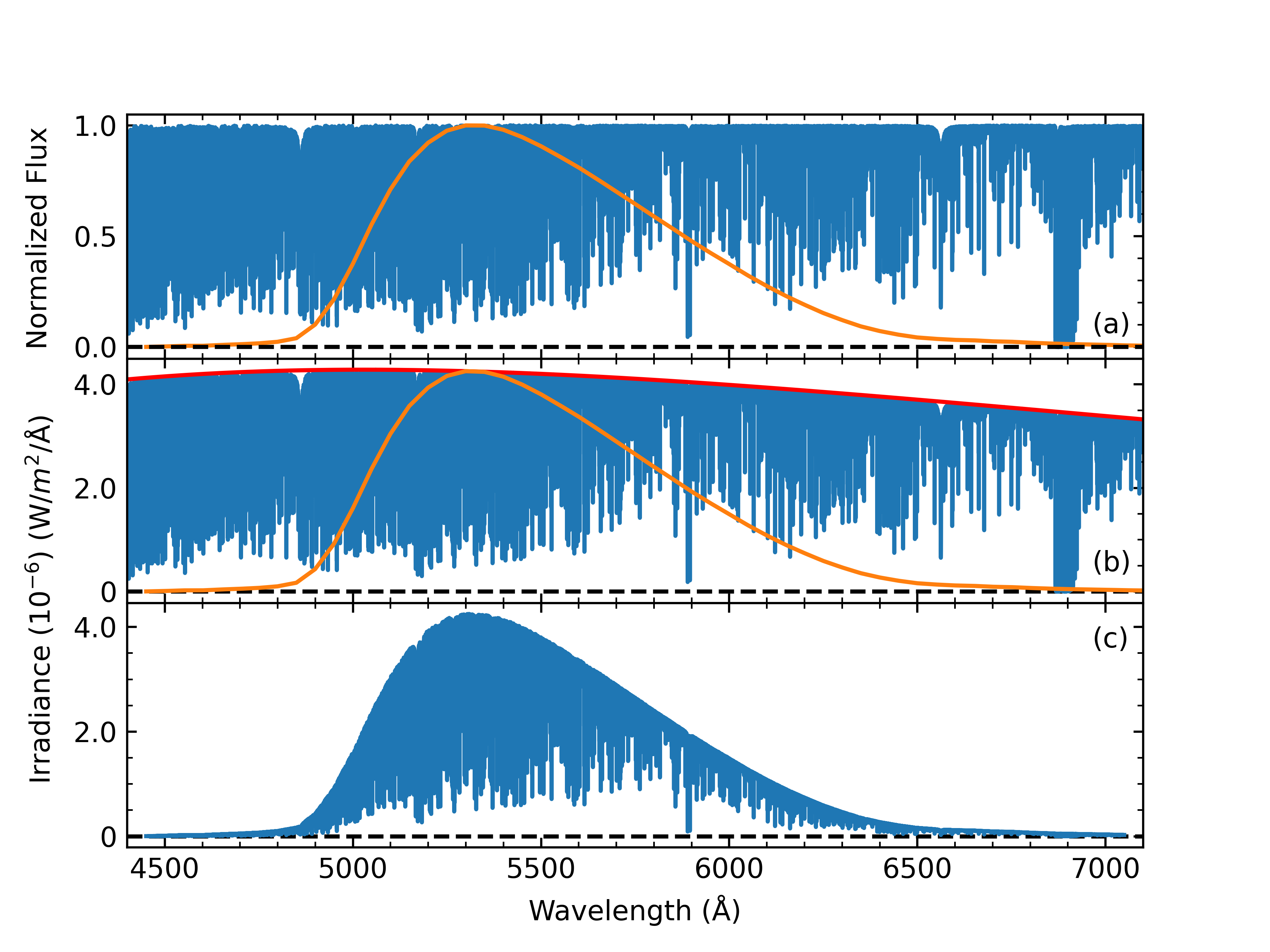}
    \caption{Normalized solar spectrum and {\it V} filter profile (a), de-normalized flux and {\it V} filter spectra (b), and the convoluted spectrum of the Sun (c). Dividing the area under the convoluted spectrum by $\sigma T_{\rm eff}^4 (\odot)$ gives solar $L_{\rm V}/L$ = 10.49 or 10.19 percent for \citet{Bessell1990} and the \citet{Landolt1992} profile functions, respectively. }  
    
    \label{fig:fv_old}
\end{figure*} 

The first step to obtain the spectroscopic $L_{\rm V}/L$ value of a star is to remove the effects of interstellar extinction on the observed spectrum. This is done by a normalization process where the continuum must equal to unity. A solar spectrum normalized to one is shown in Figure \ref{fig:fv_old}a, where the transparency profile of the visual filter normalized to the continuum is also shown. The second step is the de-normalization of the star's spectrum and the filter profile. This is done by multiplying both the Star's spectrum and filter profile by the Planck function. Figure \ref{fig:fv_old}b shows the de-normalized flux and the filter spectra, where the solar continuum and the filter profile are shown by the solid red lines. The third step is the application of the convolution process expressed by \(F_{\rm V}=\int_{0}^{\infty}S_\lambda(V)F_\lambda d\lambda\) in Equation (\ref{equ:5}). This is done by pixel-to-pixel multiplication of the de-normalized star spectrum and the filter profile. Figure \ref{fig:fv_old}c shows the convoluted solar spectrum representing the visual signal from the Sun.

Definition of the effective temperature requires that the area under the de-normalized flux spectrum must equal to $\sigma T_{\rm eff}^4$ if it were extended from zero to infinite wavelengths. However, since filtered signals correspond to a limited spectral range, one of the numerical integration techniques could be used to determine areas under the convoluted spectra. Simpson’s integration rule, via \texttt{scipy.integrate.simpson}, was adopted for this study. Finally, spectroscopic $L_{\rm V}/L$ values were obtained after dividing the convoluted areas by  $\sigma T_{\rm eff}^4$ values of the sample stars and recorded in Table \ref{tab:4} in percent units.
 
Because we have used two $S_\lambda(V)$ profile functions—one from \citet{Bessell1990} and the other from \citet{Landolt1992}, there are two columns for the spectroscopic values of $L_{\rm V}/L$ representing each of the profile functions, and the last column of Table \ref{tab:4} indicates the same relative uncertainty for both also in percent units.   

\subsection{Data to fix zero-point of $BC_V$ scale}

The value of spectroscopic $L_{\rm V}/L$ from an observed spectrum of a star is not sufficient to calculate its spectroscopic $BC_{\rm V}$. IAU2015GARB2 did not fix the zero-point of the $BC_{\rm V}$ scale. IAU2015GARB2 fixed the zero-points of absolute and apparent bolometric magnitudes by assigning definite values to the zero-point constants $C_{\rm Bol}$ and $c_{\rm Bol}$ respectively.

Therefore, the primary aim of this study is to fix the zero-point of the $BC_{\rm V}$ scale first by the help of the observational parameters of the sample stars and then to calculate individual spectroscopic $BC_{\rm V}$ of each star from their spectroscopic $L_{\rm V}/L$. How to fix the zero-point of the $BC_{\rm V}$ scale using data from a sufficient number of stars is described in detail in the Section~\ref{sec:Appendix}.

In addition to the spectroscopic $L_{\rm V}/L$ values from Table \ref{tab:4}, classically determined $BC_{\rm V}$ values from the absolute bolometric ($M_{\rm Bol}$) and visual magnitudes ($M_{\rm V}$) of stars from Tables \ref{tab:2} and \ref{tab:3} are needed to fix the zero-point constants, $C_{\rm V}$ and $c_{\rm V}$ for absolute and apparent visual magnitudes, respectively.

Classically determined $BC_{\rm V}$ values and logarithmic  $2.5 \times \log{(L_{\rm V}/L)}$ quantities for the two different profile functions are listed in Table \ref{tab:5} together with their propagated errors. Estimated individual zero point values for the $BC_{\rm V}$ scale ($C_{\rm 2}$) according to Equation (\ref{equ:5}) and propagated errors according to Equation (\ref{equ:14}) are also given for both of the profile functions \citet{Bessell1990} and \citet{Landolt1992}.

%-----------------------------------------------------------------
%Tablo 4
\begin{longtable*}{lllcccccc}

    \caption{Calculated \(L_{\rm V}/L\) values for each sample stars for each \(S_\lambda(V)\) profile with their uncertainty.}\\
    \hline
    Order & Star & Instrument &   Resolving Power & Wavelength Coverage & \textit{S/N} & $\left(L_{\rm V}/L\right)^{\textit{l}}$ & $\left(L_{\rm V}/L\right)^{\textit{b}}$ & err \\ 
         & &  & & (\AA) & & (\%) & (\%) & (\%)\\ \hline
        \endfirsthead
    \multicolumn{9}{c}%
{{\bfseries \tablename\ \thetable{}} -- continued from previous page} \\ 
\hline  
    Order & Star & Instrument &   Resolving Power & Wavelength Coverage & \textit{S/N} & $\left(L_{\rm V}/L\right)^{\textit{l}}$ & $\left(L_{\rm V}/L\right)^{\textit{b}}$ & err \\ 
    & &  & & (\AA) &  & (\%) & (\%) & (\%)\\ \hline
    \endhead
\hline \multicolumn{9}{c}{{Continued on next page}} \\ \hline
\endfoot
\hline 
\endlastfoot
    \hline
        1 &    Sun & FTS & 350\,000-500\,000 &  2\,960-13\,000 & 1349 & 10.19 & 10.49 & 0.10 \\
        2 &    HD1279    & HERMES & 85\,000 &  3\,800-9\,000  & 144 & 6.12 & 6.12 & 0.98 \\ 
        3 &    HD1404    & HERMES & 85\,000 &  3\,800-9\,000  & 209 & 10.31 & 10.40 & 0.68 \\ 
        4 &    HD1439    & HERMES & 85\,000 &  3\,800-9\,000  & 191 & 9.52 & 9.59 & 0.74 \\ 
        5 &    HD2729    & HERMES & 85\,000 &  3\,800-9\,000  & 240 & 5.54 & 5.54 & 0.59 \\ 
         ... & ... & ... & ... & ... & ... & ... & ... & ... \\
        124 &    HD220009    & PEPSI & 220\,000 &  3\,830-9\,120  & 992 & 6.27 & 6.61 & 0.14 \\ 
        125 &    HD220825    & ESPaDOnS & 68\,000 &  3\,700-10\,500  & 228 & 8.60 & 8.66 & 0.62 \\ 
        126 &    HD222173    & HERMES & 85\,000 &  3\,800-9\,000  & 223 & 7.40 & 7.42 & 0.63 \\ 
        127 &    HD222661    & HERMES & 85\,000 &  3\,800-9\,000  & 242 & 8.06 & 8.09 & 0.58 \\ 
        128 &    HD222762    & HERMES & 85\,000 &  3\,800-9\,000  & 222 & 6.52 & 6.53 & 0.64 \\
        \hline
        \multicolumn{9}{l}{$^l$, \cite{Landolt1992}; $^b$, \cite{Bessell1990}}\label{tab:4}
\end{longtable*}

%------------------------------------------------------------------

\section{Results and Discussions} \label{sec:Result}

\subsection{The Zero-Point constant of $BC$ for visual magnitudes, $C_{\rm 2}$} \label{subsec:4.1}

\begin{figure}
    \centering
    \includegraphics[width=\linewidth]{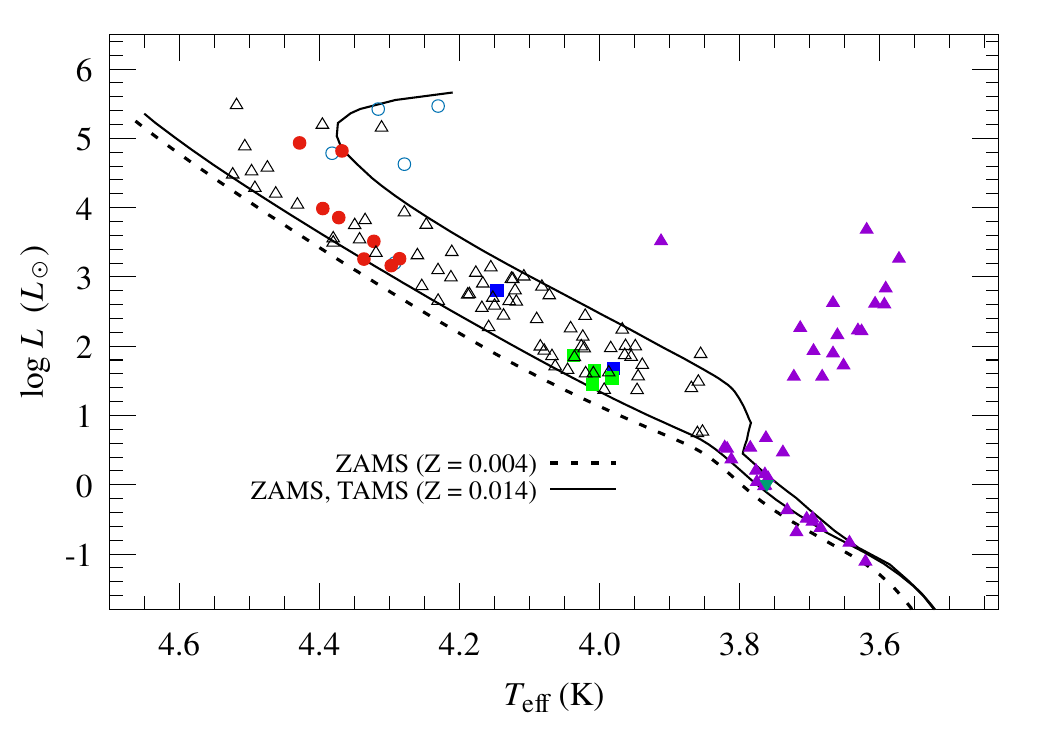}
    \caption{Distribution of the sample stars on H-R diagram. ZAMS and TAMS, according to PARSEC evolution models \citep{Bressan2012} are indicated. Symbols: $\triangle$, HERMES; \textcolor{violet}{$\filledtriangleup$}, PEPSI; $\bullet$, FIES; \textcolor{blue}{$\circ$}, FEROS; \textcolor{olive}{$\filledsquare$}, ESPaDOnS; \textcolor{blue}{$\filledsquare$}, NARVAL; \textcolor{teal}{$\filledtriangledown$}, FTS.}
    \label{fig:hr}
\end{figure}

The luminosities ($L$) listed in column 9 of Table \ref{tab:2}, from $T_{\rm eff}$ and $R$ in the same table, are presented in the form of a Hertzsprung–Russell (H–R) diagram as shown in Figure \ref{fig:hr}. In this diagram, the symbols indicate the instrument sources (Table \ref{tab:1}) from which the effective temperatures ($T_{\rm eff}$) were adopted from the literature, where model atmosphere fitting methods were employed. The $L$ of the sample stars appears distributed mostly in the main sequence, and a small fraction of them ($\sim18\%$) are stars already evolved off the main sequence, but not up to white dwarfs. Thus, this distribution fulfilled our a priori condition in the first approximation that the zero-point constants ($C_{\rm 2}$, $C_{\rm V}$, $c_{\rm V}$) to be determined by this study should not be biased by the position of stars on the H-R diagram.  

%-----------------------------------------------------------------------------------------------------
%Figure 3

%-----------------------------------------------------------------------------------------------------
%Figure 4
\begin{figure}
    \centering
    \includegraphics[width=\linewidth]{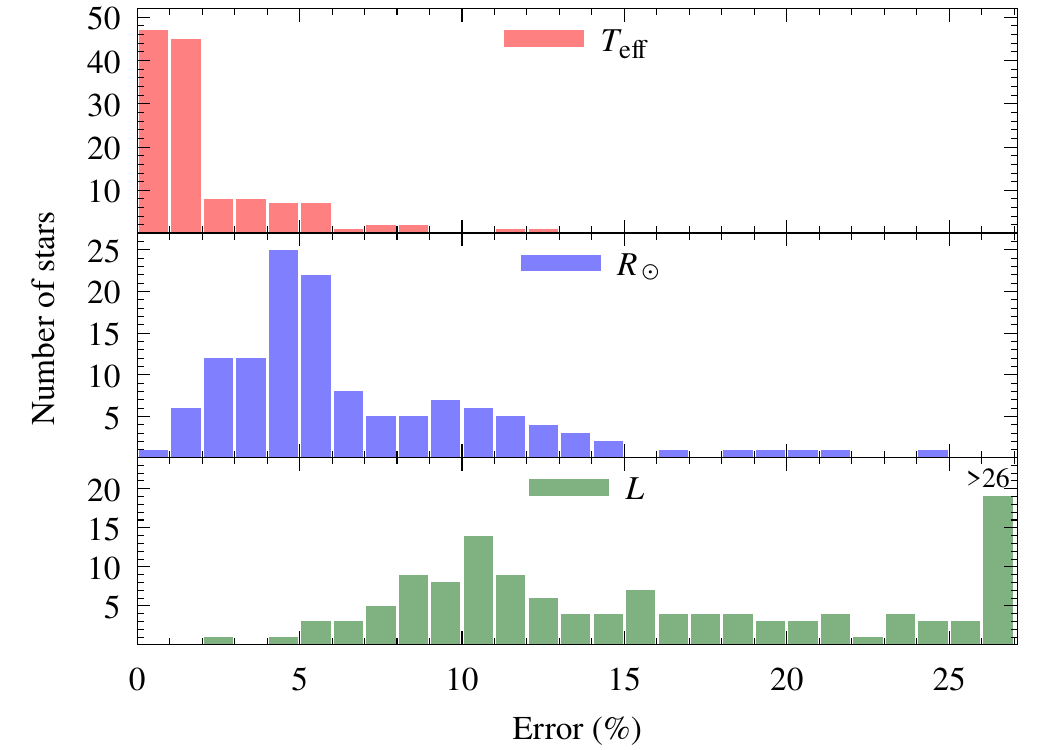}
    \caption{Uncertainty histograms of $T_{\rm eff}$ (top), $R$ (middle), and $L$ (bottom).}
    \label{fig:t_r_l_uncertainty}
\end{figure}
%-----------------------------------------------------------------------------------------------------

Figure \ref{fig:hr} also shows that the sample stars are distributed in effective temperatures from 3\,779 K (HD\,18884) to 33\,400 K (HD\,36512), and the sizes (radii) are from 0.53 $R_\odot$ (HD\,201092) to 133.53 $R_\odot$ (HD\,186791). Uncertainty contributions of the stellar parameters in the computed $L$ are given in Figure \ref{fig:t_r_l_uncertainty}, where the typical observational uncertainty of a $T_{\rm eff}$ is 1\%, while the typical uncertainty of a radius is about 5\%. Because $L$ is proportional to the square of $R$ and the fourth power of $T_{\rm eff}$, the uncertainty of $L$ (bottom of Figure \ref{fig:t_r_l_uncertainty}) inflates to larger values where the typical error of $L$ is about 10\% which confirms \cite{Eker2021b}, who was previously estimated the typical uncertainty as a range 8.2\% – 12.2\% recently, but not a single value.   

%----------------------------------------------------------------------------------------------------
%Figure 5

\begin{figure}
    \centering
    \includegraphics[width=\linewidth]{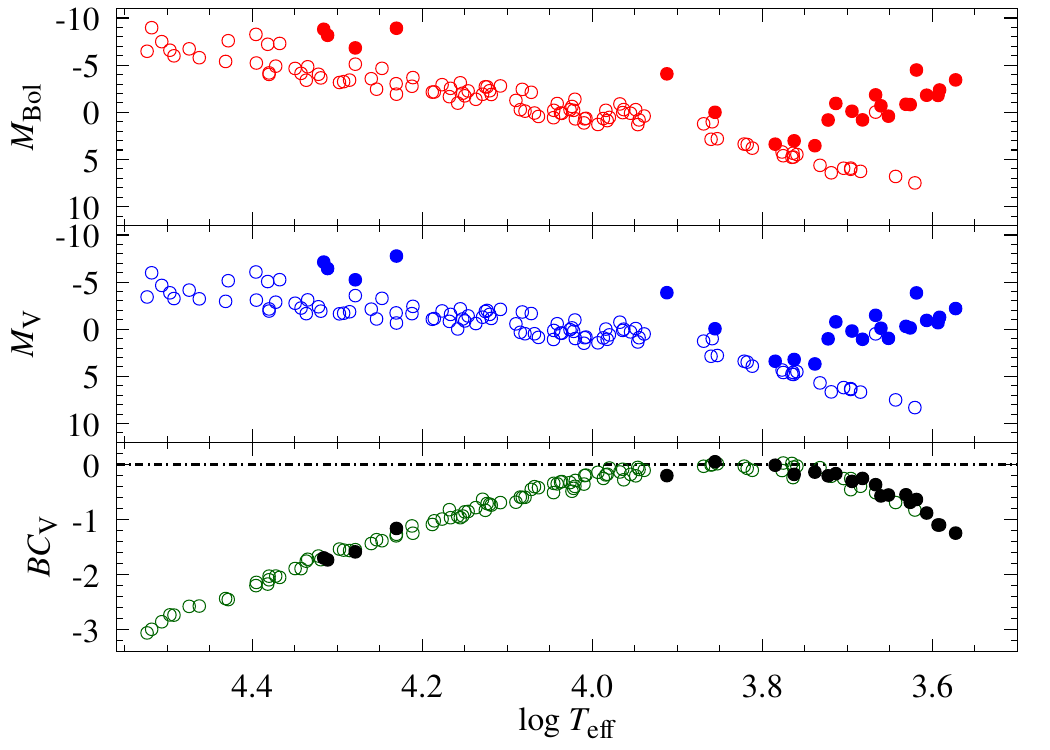}
    \caption{Bolometric correction ($BC_{\rm V}$) for the visual magnitudes (bottom) as the difference between the absolute bolometric (top) and the visual (middle) magnitudes. Evolved stars are shown by the filled symbols.}
    \label{fig:mbol_mv_bc}
\end{figure}

%----------------------------------------------------------------------------------------------------

Distributions of the computed absolute bolometric ($M_{\rm Bol}$), and visual ($M_{\rm V}$) magnitudes against $\log T_{\rm eff}$ are shown in Figure \ref{fig:mbol_mv_bc} together with the resulting bolometric corrections ($BC_{\rm V}$) according to Equation~(\ref{equ:1}). The distributions in Figure~\ref{fig:hr} and in the top two panels of Figure~\ref{fig:mbol_mv_bc} are all called H-R diagrams. One normally would expect a similar distribution because of the common name: H-R diagram. However, in this study, we have observed that the choice of the vertical axis can lead to slight variations in the apparent shapes and distributions. The smooth concave curvature of the main-sequence stars is more noticeable in the two panels in Figure~\ref{fig:mbol_mv_bc} than the curvature in Figure~\ref{fig:hr}. With a careful look, the curvature in the middle box is a little stronger than the curvature in the top box in Figure~\ref{fig:mbol_mv_bc}. Apparently, the shape of the $BC_{\rm V}-T_{\rm eff}$ relation in the bottom panel is governed by the difference in the concavity of the two distributions shown above it. Although the evolved stars are not located within the band of the main-sequence stars, and despite the difference is much bigger towards the cooler end, the $BC_{\rm V}-T_{\rm eff}$ curve of the evolved stars (filled black circles) seems to follow the same $BC_{\rm V}-T_{\rm eff}$ relation as the main-sequence stars. 

\begin{figure}
    \centering
    \includegraphics[width=\linewidth]{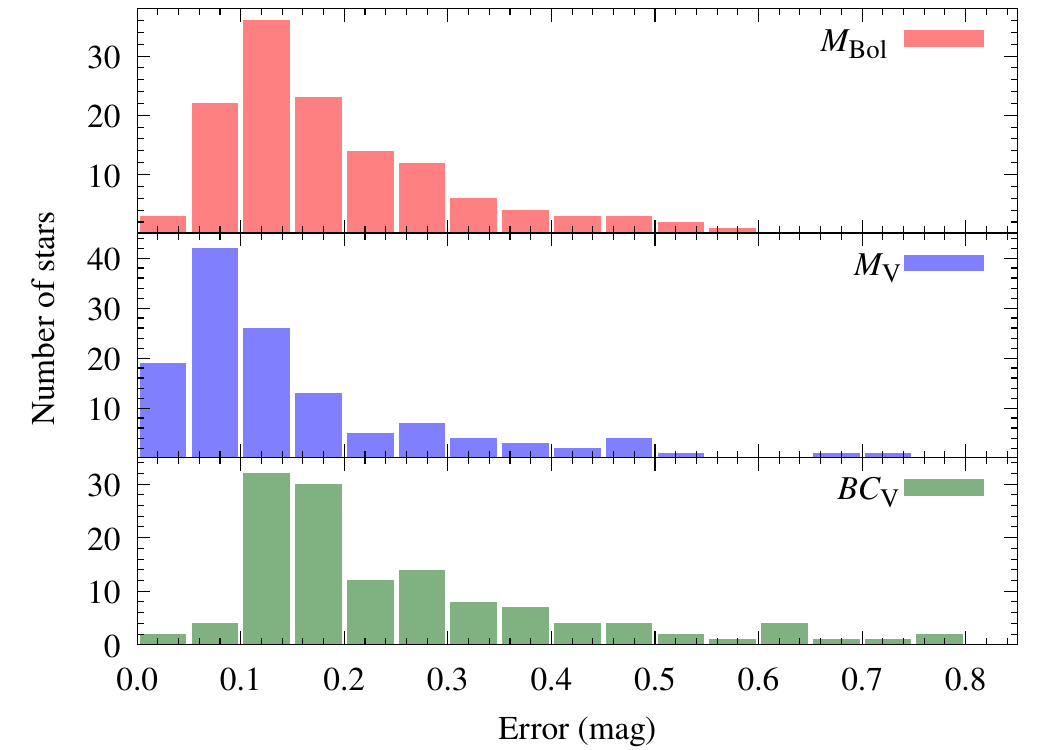}
    \caption{The uncertainty histograms of $M_{\rm Bol}$ (top), $M_{\rm V}$ (middle), and $BC_{\rm V}$ (bottom)}
    \label{fig:mbol_mv_bc_uncertainty}
\end{figure}

Error distributions of $M_{\rm Bol}$, $M_{\rm V}$ and $BC_{\rm V}$ are shown in Figure~\ref{fig:mbol_mv_bc_uncertainty}. The errors of $M_{\rm Bol}$ are the propagated errors of $L$ according to Equation~(\ref{equ:3}), while the errors of $M_{\rm V}$ are the propagated errors from the uncertainties of the observed parameters; apparent magnitudes ($V$), trigonometric parallaxes ($\varpi$) and interstellar extinctions ($A_{\rm V}$) (see Table~\ref{tab:3}). 

%----------------------------------------------------------------------------------------------------
%Figure 6
%----------------------------------------------------------------------------------------------------

\begin{figure}
    \centering
    \includegraphics[width=\linewidth]{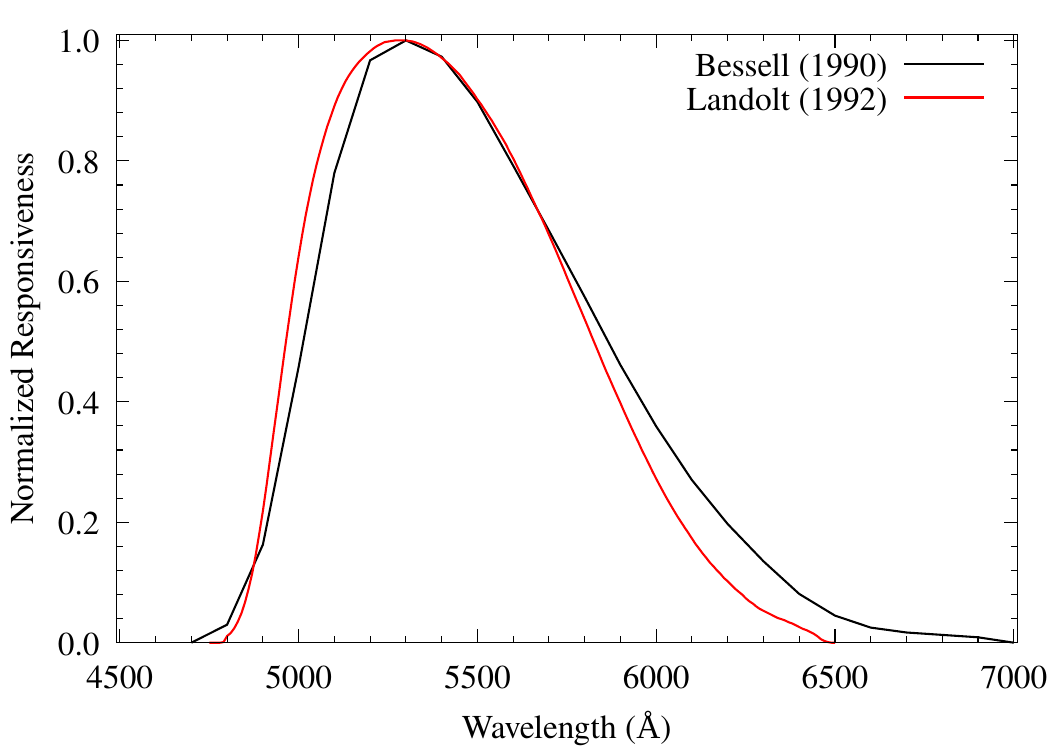}
    \caption{Normalized transparency profiles of the $V$ filter by \cite{Bessell1990} and \cite{Landolt1992}.}
    \label{fig:filters}
\end{figure}

Visual to bolometric flux ratio or fraction of the bolometric flux $\left(L_{\rm V}/L\right)$ through the visual filter is the very basic parameter that is measured from a spectrum of a star for calculating its spectroscopic $BC_{\rm V}$ according to Equation~(\ref{equ:18}). The shape of the wavelength profile of the visual filter $\left(S_\lambda(V)\right)$ is very critical not only for obtaining $\left(L_{\rm V}/L\right)$ from an observed spectrum but also for determining the value of the zero-point constant $C_{\rm 2}$ for a star. Various authors used various transparency profiles, which are all listed in The Asiago Photometric Database\footnote{\url{http://ulisse.pd.astro.it/Astro/ADPS/}} \citep{Moro2000, Fiorucci2003}, all representing the visual filter. We have examined them all and decided to use the visual profile functions of \cite{Bessell1990} and \cite{Landolt1992} that appear to be of slightly different shapes (see Figure~\ref{fig:filters}). It is important to know which filter shape best represents the $V$ magnitudes collected from the SIMBAD database. Because both were equally likely to be used in the measurements by photomultiplier tubes in the past or perhaps CCD observations in the near past, which now appear to be listed in the SIMBAD database, we have decided to use both to understand and show how the filter profile changes the values of $\left(L_{\rm V}/L\right)$ and consequently $C_{\rm 2}$ values at last.   
%----------------------------------------------------------------------------------------------------
%Figure 7

%----------------------------------------------------------------------------------------------------
%Figure 8
\begin{figure}
    \centering
    \includegraphics[width=\linewidth]{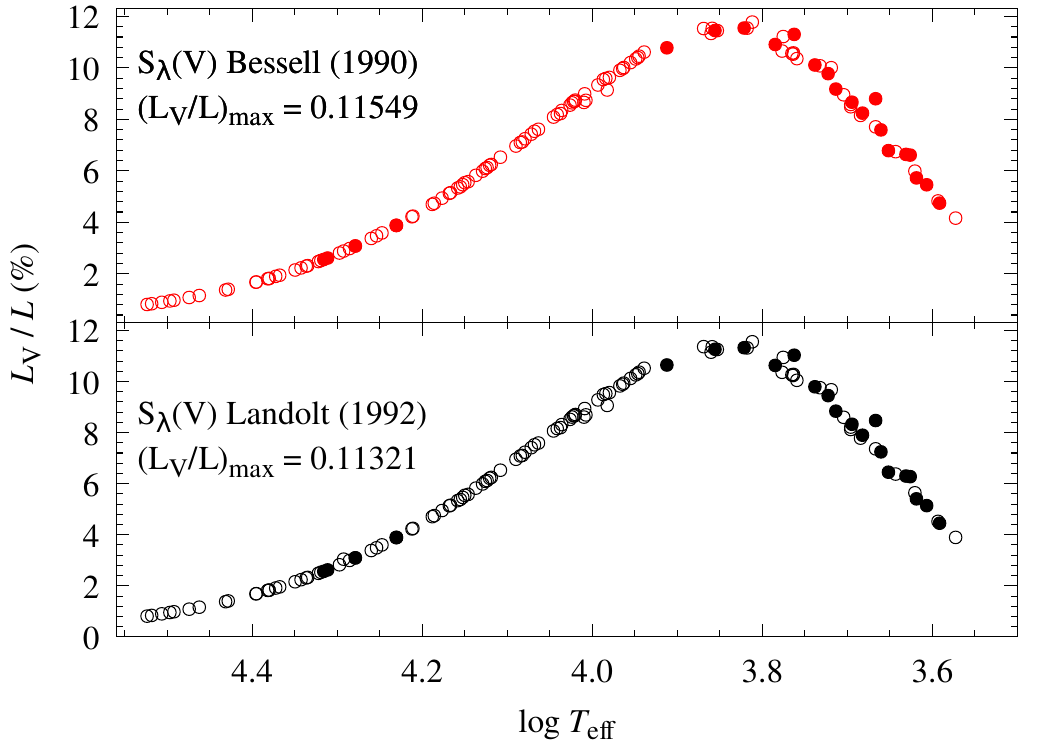}
    \includegraphics[width=\linewidth]{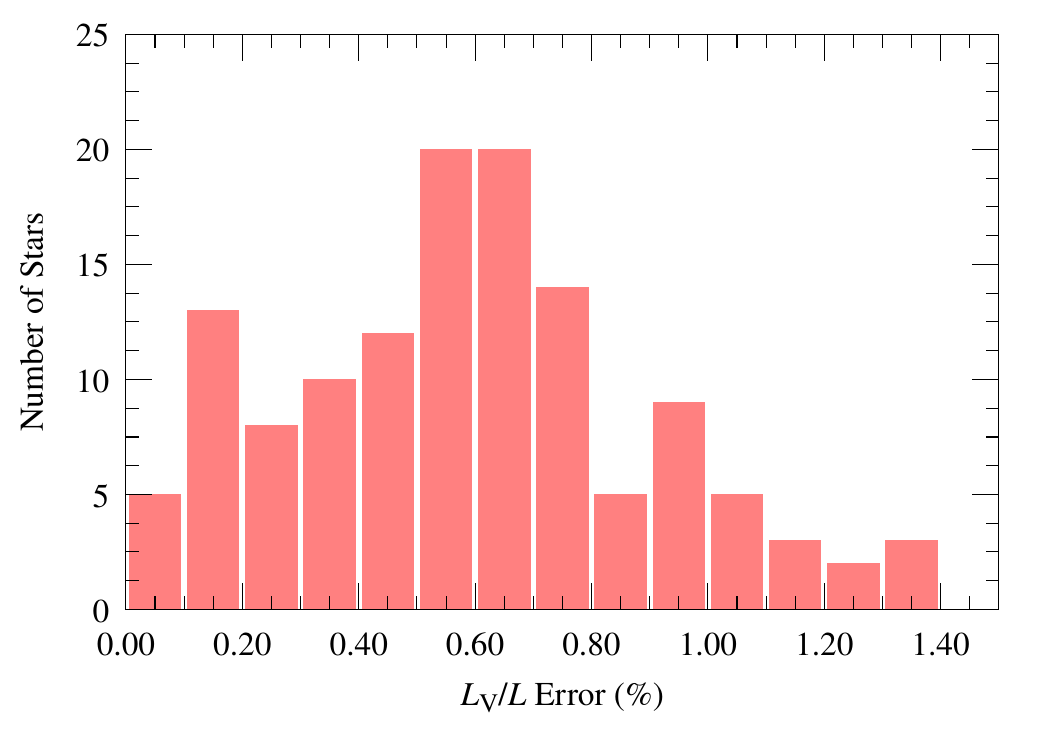}
    \caption{Visual to bolometric ratio, or fraction of the bolometric flux $\left(L_{\rm V}/L\right)$ according to functions from \cite{Bessell1990} and \cite{Landolt1992} indicated (above).  Relative errors of $\left(L_{\rm V}/L\right)$ (below).}
    \label{fig:lv_l_teff}
\end{figure}
%----------------------------------------------------------------------------------------------------

Variation of the visual to bolometric flux ratio $\left(L_{\rm V}/L\right)$ for a hypothetical star across the range of effective temperatures in an H-R diagram are demonstrated in Figure~\ref{fig:lv_l_teff}, where the empirically predicted shapes as a function of $T_{\rm eff}$ appears to be very similar despite the $V$ filter profiles of \cite{Bessell1990} and \cite{Landolt1992} showing noticeably different shapes in Figure~\ref{fig:filters}. We have recorded both of the maximum values in the two boxes in Figure~\ref{fig:lv_l_teff} in order to show a very small (0.2\%) systematic difference between $\left(L_{\rm V}/L\right)$ values computed by the two $\left(S_\lambda(V)\right)$ profile functions noticeable in the third digit after the decimal. The systematic difference appears to be slightly decreasing towards the hotter end and vice versa, slightly increasing towards the cooler end. The relative uncertainty of an empirical $\left(L_{\rm V}/L\right)$ value is estimated by realizing that the integration and truncation errors are too small, thus, ignored, while assuming the uncertainties of visual and bolometric signals are the same and both are characterized by the \textit{S/N} ratio of the spectrum concerned. The distribution of the relative uncertainties in the visual to bolometric flux ratio $\Delta \left(L_{\rm V}/L\right) / \left(L_{\rm V}/L\right)$ is displayed as a histogram format in Figure~\ref{fig:lv_l_teff}, where the typical uncertainty is 0.6\%.

As shown in Figure~\ref{fig:lv_l_teff}, the uncertainties in   $L_{\rm V}/L$ appear much smaller than the uncertainties in $BC_{\rm V}$, which were shown in Figure~\ref{fig:mbol_mv_bc_uncertainty}. This should be the result of the high \textit{S/N} of the sample spectra. Also, the shape of the visual to bolometric flux ratio is noticeably different than the shape of the $BC_{\rm V}-T_{\rm eff}$ relation given in Figure~\ref{fig:mbol_mv_bc}. This is because the $BC_{\rm V}$ values are in magnitude scale, while the $\left(L_{\rm V}/L\right)$ values are just simple ratios. If logarithms of the $\left(L_{\rm V}/L\right)$ were taken and then multiplied by 2.5, the shape of the $2.5 \log{\left(L_{\rm V}/L\right)}-T_{\rm eff}$ function would have been very similar to the shape of the $BC_{\rm V}-T_{\rm eff}$ function. This is because the star-by-star difference of these two functions defines the value of the zero-point constant $C_{\rm 2}$, which is just a numerical constant same for all stars.

%------------------------------------------------
%Tablo 5
\begin{longtable*}{llcccccccccc}

    \caption{Photometric and spectroscopic data to calculate $C_2$ the zero-point constant of $BC$ for the visual magnitudes.}\\
    \hline
             & & & & \multicolumn{4}{c}{$S_{\lambda}$(V) from \citet{Landolt1992}} & \multicolumn{4}{c}{$S_{\lambda}$(V) from \citet{Bessell1990}} \\
    Order & Star & $BC$ & err &  $2.5\times\log\frac{L_{\rm V}}{L}$ &  err & $C_{\rm 2}$ & err &  $2.5\times\log\frac{L_{\rm V}}{L}$ &  err & $C_{\rm 2}$ & err\\
         & & \multicolumn{2}{c}{(mag)} &  \multicolumn{2}{c}{(mag)} & \multicolumn{2}{c}{(mag)}& \multicolumn{2}{c}{(mag)}& \multicolumn{2}{c}{(mag)} \\
\hline
        \endfirsthead
    \multicolumn{12}{c}%
{{\bfseries \tablename\ \thetable{}} -- continued from previous page} \\ 
\hline  
             & & & & \multicolumn{4}{c}{$S_{\lambda}$(V) from \citet{Landolt1992}} & \multicolumn{4}{c}{$S_{\lambda}$(V) from\citet{Bessell1990}} \\
    Order & Star & $BC$ & err &  $2.5\times\log\frac{L_{\rm V}}{L}$ &  err & $C_{\rm 2}$ & err &  $2.5\times\log\frac{L_{\rm V}}{L}$ &  err & $C_{\rm 2}$ & err\\
         & & \multicolumn{2}{c}{(mag)} &  \multicolumn{2}{c}{(mag)} & \multicolumn{2}{c}{(mag)} & \multicolumn{2}{c}{(mag)} & \multicolumn{2}{c}{(mag)} \\ 
\hline
\endhead
\hline \multicolumn{12}{c}{{Continued on next page}} \\ \hline
\endfoot
\hline 
\endlastfoot
    \hline
        1 &    Sun       & -0.070 & 0.030 & -2.480 & 0.001 & 2.410 & 0.030 & -2.448 & 0.001 & 2.378 & 0.030 \\ 
        2 &    HD1279    & -0.683 & 0.154 & -3.033 & 0.011 & 2.350 & 0.155 & -3.032 & 0.011 & 2.350 & 0.155 \\ 
        3 &    HD1404    & -0.028 & 0.156 & -2.467 & 0.007 & 2.439 & 0.156 & -2.458 & 0.007 & 2.430 & 0.156 \\ 
        4 &    HD1439    & -0.163 & 0.121 & -2.553 & 0.008 & 2.390 & 0.122 & -2.546 & 0.008 & 2.383 & 0.122 \\
        5 &    HD2729    & -0.835 & 0.326 & -3.141 & 0.006 & 2.306 & 0.326 & -3.141 & 0.006 & 2.306 & 0.326 \\ 
        ... & ... & ... & ... & ... & ... & ... & ... & ... & ... & ... & ... \\
        124 &    HD220009    & -0.661 & 0.186 & -3.006 & 0.002 & 2.345 & 0.186 & -2.950 & 0.002 & 2.288 & 0.186 \\ 
        125 &    HD220825    & -0.329 & 0.175 & -2.663 & 0.007 & 2.334 & 0.175 & -2.657 & 0.007 & 2.327 & 0.175 \\ 
        126 &    HD222173    & -0.427 & 0.197 & -2.826 & 0.007 & 2.399 & 0.197 & -2.824 & 0.007 & 2.396 & 0.197 \\ 
        127 &    HD222661    & -0.485 & 0.233 & -2.734 & 0.006 & 2.249 & 0.233 & -2.730 & 0.006 & 2.245 & 0.233 \\ 
        128 &    HD222762    & -0.667 & 0.250 & -2.964 & 0.007 & 2.296 & 0.250 & -2.963 & 0.007 & 2.295 & 0.250
        \label{tab:5}
\end{longtable*}

\begin{figure}
    \centering
    \includegraphics[width=\linewidth]{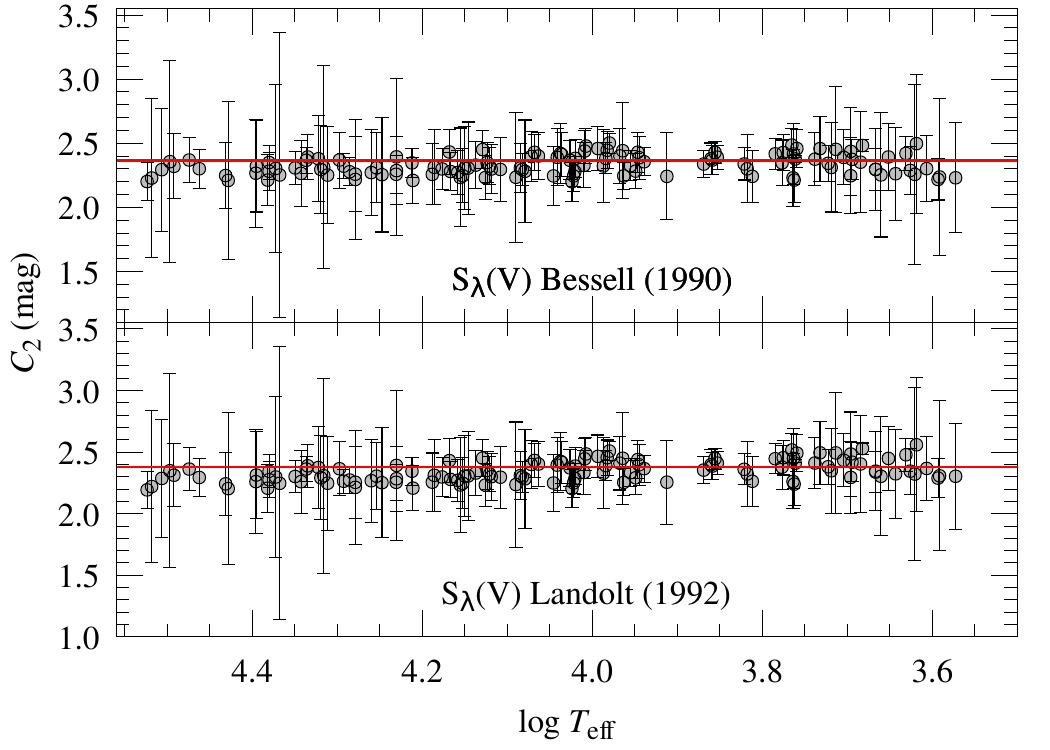}
    \caption{The zero-point constant values for the $BC_{\rm V}$ scale from the sample spectra containing 128 stars. Error bars are the propagated errors from the observational uncertainties. Horizontal lines mark the values of the weighted mean: $C_{\rm 2}= 2.3653$ If $S_\lambda(V)$ function of \cite{Bessell1990} is used, $C_{\rm 2}= 2.3826$ if $S_\lambda(V)$ function of \cite{Landolt1992} is used.}
    \label{fig:teff_C2}
\end{figure}

The two types of $C_{\rm 2}$, one for each filter, are plotted in Figure~\ref{fig:teff_C2} against effective temperatures. Because each $C_{\rm 2}$ is independent and has a propagated uncertainty, we have calculated both arithmetic/weighted means and recorded them in Table~\ref{tab:6}, where the standard deviations and standard errors are also indicated for both of the profile functions. 

%----------------------------------------------------------------------------------------------------
%Figure 9

%----------------------------------------------------------------------------------------------------
%Tablo 6
\begin{table*}
    \centering
    \caption{Statistics of the zero-point constant of the $BC_{\rm V}$ scale $C_{\rm 2}$ estimated from the sample spectra and spectroscopic ($T_{\rm eff}$), photometric ($V$), and astrometric ($\varpi$) observations.}
    \begin{tabular}{ccccccc}
    \hline
        Arithmetic Mean of $C_{\rm 2}$ & S.D. & S.E. & Weighted Mean of $C_{\rm 2}$ & S.D. & S.E.  & Source of $S_\lambda(V)$\\
        \hline
         2.3293 & 0.0768 & 0.007 & 2.3653 & 0.0756 & 0.0067 & \cite{Bessell1990} \\
         2.3423 & 0.0832 & 0.007 & 2.3826 & 0.0856 & 0.0076 & \cite{Landolt1992}\\
         \hline
    \end{tabular}
    \label{tab:6}
\end{table*}
%----------------------------------------------------------------------------------------------------

The transparency profile of the filter $S_\lambda(V)$ used in photometric observations must be the same as the one used when extracting the visual to bolometric ratio $\left(L_{\rm V}/L\right)$ from a spectrum of a star when determining the value of $C_{\rm 2}$. Ignoring this detail and using another $S_\lambda(V)$ profile, other than the one given in Table~\ref{tab:6}, will result in differences of no more than 0.02 mag from the values listed there. Using the same profile function is important to achieve an accuracy of 7-8 millimagnitude level, when computing $BC_{\rm V}$ of a star from its spectrum by the method described in this study, which requires only a normalized observed spectrum and an accurately determined $T_{\rm eff}$ and $S_\lambda(V)$ unlike classically computed standard $BC$  requiring $R$, $V$, $\varpi$ and $A_{\rm V}$ in addition to $T_{\rm eff}$.    

\subsection{Spectroscopic BC and $BC - T_{\rm eff}$ relation of visual magnitudes}

A spectroscopic $BC$ from a reliable spectrum is standard by definition. Provided with an accurate trigonometric parallax ($\varpi$) and an interstellar extinction ($A_{\rm V}$), standard $BC_{\rm V}$ would be a very useful parameter to obtain the standard $L$ of a star from its apparent visual magnitude ($V$). 

If a high-resolution high \textit{S/N} ratio spectrum of the star covering wavelength range of the visual filter $\left(S_\lambda(V)\right)$ is not available, then a reliable standard $BC_{\rm V}-T_{\rm eff}$ relation would be the only way to get a standard $BC_{\rm V}$ for the same purpose. In addition to tabulated tables of $BC_{\rm V}$, analytical relations in the form of fifth, fourth, and third degree polynomials representing $BC_{\rm V}-T_{\rm eff}$ relations at three temperature regimes were first introduced to astrophysics by \cite{Flower1996}. The empirically determined coefficients of these functions were rectified later by \cite{Torres2010} before an updated $BC_{\rm V}-T_{\rm eff}$ relation derived empirically from the astrophysical parameters of Detached Double-Lined Eclipsing Binaries \citep[DDEB,][]{Eker2014} is announced by \cite{Eker2020}. The empirical multi-band (Johnson $B$, $V$ and Gaia $G$, $G_{\rm BP}$, $G_{\rm RP}$) standard $BC-T_{\rm eff}$ relations were fixed later by \cite{Bakis2022} for further increasing the accuracy of predicted $L$ of single stars.

In this study, we introduce another new concept, the spectroscopic $BC_{\rm V}-T_{\rm eff}$ relation in addition to the spectroscopic $BC_{\rm V}$. Using the filter profile function $\left(S_\lambda(V)\right)$ of \cite{Bessell1990} for the one and the profile function of \cite{Landolt1992} for the other, the two spectroscopic relations $BC_{\rm V}-T_{\rm eff}$ were calibrated by fitting fourth-degree polynomials according to the least-squares method to the $BC_{\rm V}$ data obtained from the high-resolution high \textit{S/N} spectra of 128 stars. We also evaluated the classical $BC_{\rm V}$ data (Figure~\ref{fig:mbol_mv_bc}) by fixing a photometric $BC_{\rm V}-T_{\rm eff}$ relation by the same method, also represented by a fourth-degree polynomial. Coefficients, errors in the coefficients, and related statistics (standard deviations and correlations), validity ranges, the two temperatures at which $BC_{\rm V}=0$ mag, the temperatures corresponding to the maximum and $BC_{\rm V}$ at the solar effective temperature (5\,772 K) are all given in Table~\ref{tab:7} for the three functions separately.

%----------------------------------------------------------------------------------------------
%Tabo 7
\begin{table*}
\centering
    \caption{$BC_{\rm V}-T_{\rm eff}$ functions determined by fitting a fourth-degree polynomial according to the least-squares method to the spectroscopic (upper and middle) and photometric (lower) $BC_{\rm V}$ and $T_{\rm eff}$ data.}
   \begin{tabular}{ccccccc}
    \toprule
        & \multicolumn{5}{c}{$BC = a + b \times (\log{T_{\rm eff}}) + c \times (\log{T_{\rm eff}})^2 + d \times (\log{T_{\rm eff}})^3 + e \times (\log{T_{\rm eff}})^4$} \\
        & $a$ & $b$ & $c$ & $d$ & $e$ \\
        \hline
        \parbox[t]{2mm}{\multirow{7}{*}{\rotatebox[origin=c]{90}{\citet{Bessell1990}}}} & -2120.5827 & 1941.0005 & -665.5339 & 101.5261 & -5.8311 \\
        & ($\pm$135.9673) & ($\pm$134.8772) & ($\pm$50.0903) & ($\pm$8.2541) & ($\pm$0.5092) \\
        & \multicolumn{5}{c}{$\sigma = 0.023$, $R^2 = 0.9991$}\\
        & \multicolumn{5}{c}{valid in the range $3\,738 \leq T_{\rm eff} \leq 33\,400$ K}\\
        & & $BC = 0.00$ & $T_{\rm eff,1}= 6\,518$ K & $T_{\rm eff,2}= 7\,642$ K \\
        & & $BC_{\rm max} = 0.014$& $T_{\rm eff}= 7\,049$ K\\
        & & $BC_\odot = -0.081$ & $T_\odot = 5\,772$ K  \\
                \hline
        \parbox[t]{2mm}{\multirow{7}{*}{\rotatebox[origin=c]{90}{\citet{Landolt1992}}}} & -2048.5449 & 1873.0893  & -641.4593 & 97.7231 & -5.6053 \\
        & ($\pm$125.5375) & ($\pm$124.5310) & ($\pm$46.2479) & ($\pm$7.6209) & ($\pm$0.4702) \\
        & \multicolumn{5}{c}{$\sigma = 0.021$, $R^2 = 0.9992$}\\
        & \multicolumn{5}{c}{valid in the range $3\,738 \leq T_{\rm eff} \leq 33\,400$ K}\\
        & & $BC = 0.00$ & $T_{\rm eff,1}= 6\,400$ K & $T_{\rm eff,2}= 7\,611$ K \\
        & & $BC_{\rm max} = 0.016$& $T_{\rm eff}= 6\,967$ K\\
        & & $BC_\odot = -0.067$ & $T_\odot = 5\,772$ K  \\
                \hline

        \parbox[t]{2mm}{\multirow{7}{*}{\rotatebox[origin=c]{90}{Photometric}}} & -2654.6516 & 2454.2763 & -849.9471 & 130.8906 & -7.5802  \\
        & ($\pm$403.9654) & ($\pm$400.7265) & ($\pm$148.8205) & ($\pm$24.5233) & ($\pm$1.5129) \\
        & \multicolumn{5}{c}{$\sigma = 0.068$, $R^2 = 0.9927$}\\
        & \multicolumn{5}{c}{valid in the range $3\,738 \leq T_{\rm eff} \leq 33\,400$ K}\\
        & & $BC = 0.00$ & $T_{\rm eff,1}= 6\,194$ K & $T_{\rm eff,2}= 7\,775$ K \\
        & & $BC_{\rm max} = 0.031$& $T_{\rm eff}= 6\,918$ K\\
        & & $BC_\odot = -0.055$ & $T_\odot = 5\,772$ K  \\
        \bottomrule
    \end{tabular}
    \label{tab:7}
\end{table*}

%----------------------------------------------------------------------------------------------
%Figure 10
\begin{figure}[b]
    \centering
    \includegraphics[width=\linewidth]{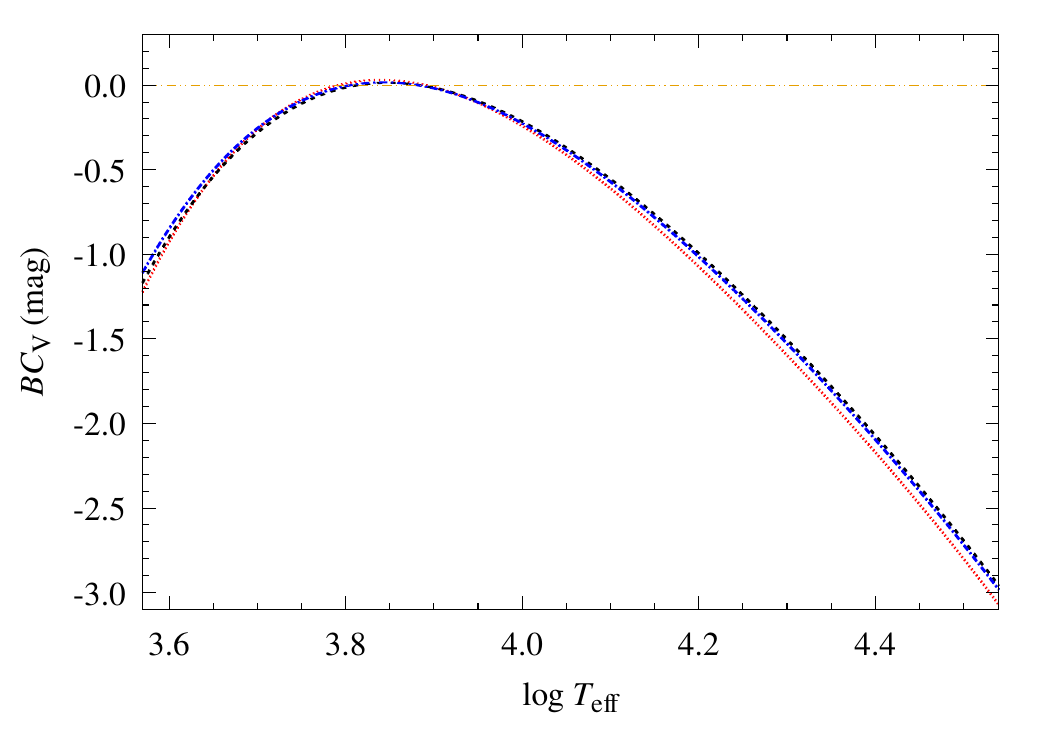}
    \caption{Curves of the analytical $BC_{\rm V}-T_{\rm eff}$ relations in Table~\ref{tab:7}. Black dashed curve uses $S_\lambda(V)$ from \cite{Bessell1990}, Blue dashed-dotted curve uses $S_\lambda(V)$ from \cite{Landolt1992}, red dotted curve is the photometric $BC_{\rm V}-T_{\rm eff}$ curve.}
    \label{fig:fitler}
\end{figure}
%----------------------------------------------------------------------------------------------

Analytical curves of the three (two spectroscopic, one photometric) functions are compared in Figure~\ref{fig:fitler}. All cross the horizontal axis and thus have $BC_{\rm V}=0$ mag at the temperatures as indicated in the Table~\ref{tab:7}. All show a very similar shape. Nevertheless, the accuracies of the spectroscopic relations are better than the photometric relation, as indicated by their standard deviations almost three times narrower than the standard deviation of the photometric $BC_{\rm V}-T_{\rm eff}$ relation.

Both spectroscopic functions are equally likely to give the most accurate $BC_{\rm V}$ for a star from an analytical relation. The relation associated with $S_\lambda(V)$ from \cite{Bessell1990} indicates that the Sun has visual absolute brightness $M_{\rm V,\odot}= M_{\rm Bol,\odot} - BC_{\rm V} =4.74 + 0.081 = 4.821 ±0.023$ mag, while the $BC_{\rm V,\odot} =-0.082$ from the solar spectrum directly indicates $M_{\rm V,\odot}= 4.822 ±0.001$ and $V_{\odot}= -26.750 ±0.001$ mags for the visual absolute and apparent magnitudes, respectively, for the Sun.

The relation associated with \cite{Landolt1992} indicates that the Sun has visual absolute brightness $M_{\rm V,\odot}= M_{\rm Bol,\odot} - BC_{\rm V} =4.74 + 0.067 = 4.807 ±0.021$ mag, while the $BC_{\rm V,\odot} =-0.097$ mag from the solar spectrum directly indicates $M_{\rm V,\odot}= 4.837 ±0.001$ and $V_{\odot}= -26.735 ±0.001$ mags for the visual absolute and apparent magnitudes, respectively, for the Sun. 

It is clear that a direct measurement of $BC_{\rm V}$ from an observed spectrum must be preferred rather than estimating it from a pre-calibrated relation because using a pre-calibration relation adds an extra uncertainty to the $BC_{\rm V}$ first, then it will propagate together with the other observational uncertainties in predicting the standard $L$ of the star.  

\subsection{The Zero-Point constants of visual magnitudes, $C_{\rm V}$ and $c_{\rm V}$ }

Once the zero-point constant of the $BC_{\rm V}$ scale, $C_{\rm 2}$ is determined, then it is straight forward to calculate the zero-point constants of absolute and apparent visual magnitudes, $C_{\rm V}$ and $c_{\rm V}$, using Equations~(\ref{equ:15}) and (\ref{equ:16}). The calculated values of $C_{\rm V}$ and $c_{\rm V}$ are given in Table~\ref{tab:8} where the columns and rows are self-explanatory also to display luminosities if absolute visual brightness ($M_{\rm V}$) of a star is zero and irradiances (fluxes) just above the Earth’s atmosphere if apparent visual brightness ($V$) of a star is zero at SI and cgs units not only for the $C_{\rm 2}$ value determined using $S_\lambda(V)$ profile function of \cite{Bessell1990} but also for the $C_{\rm 2}$ value determined from the $S_\lambda(V)$ profile function of \cite{Landolt1992}. A reader may prefer one of the $C_{\rm V}$ values to calculate the visual luminosity of the star directly from its absolute visual magnitude according to 
Equation (\ref{equ:4}). After this study, it is now also possible to calculate the irradiance of the visual photons, or the visual flux just above the Earth's atmosphere according to Equation (\ref{equ:17}) (if there is no interstellar extinction) from the apparent brightness ($V$) of the star after choosing one of the $c_{\rm V}$ values given in Table~\ref{tab:8}. We believe that these opportunities will open new paths for model atmosphere studies.

%----------------------------------------------------------------------------------------------
%Tablo 8
\begin{table*}
    \centering
    \caption{The values of the zero-point (ZP) constants of absolute and apparent magnitudes for bolometric and visual brightnesses: $C_{\rm Bol}$, $C_{\rm V}$, $c_{\rm Bol}$, $c_{\rm v}$. Luminosities if $M_{\rm Bol}=0$ and $M_{\rm V}=0$ mag and fluxes (irradiances) if $m_{\rm Bol}=0$ and $V=0$ mag.}
    \begin{tabular}{ccccccc}
    \toprule
    & ZP of absolute mag & if $M_{\rm Bol}=0$ & Unit & ZP of apparent mag & if $m_{\rm Bol}=0$ & Unit \\
    & $C_{\rm Bol}$ & $L_{\rm 0}$ (Bol) &  & $c_{\rm Bol}$ & $f_{\rm 0}$ (Bol) &  \\
    \hline
    SI  & 71.197425 & 3.0128E+28 & W            & -18.997351 & 2.5180E-08 & W m$^{-2}$             \\
    cgs & 88.697425 & 3.0128E+35 & erg s$^{-1}$ & -11.497351 & 2.5180E-06 & erg s$^{-1}$ cm$^{-2}$ \\
    \hline
    \multicolumn{7}{c}{$C_{\rm V} = C_{\rm Bol} -2.3653$ \qquad \qquad  $S_\lambda(V)$ from \cite{Bessell1990} \qquad \qquad $c_{\rm V} = c_{\rm Bol} -2.3653$} \\
    \
    SI  & 68.8321 & 3.4107E+27 & W              & -21.3627   & 2.8506E-09 & W m$^{-2}$             \\
    cgs & 86.3321 & 3.4107E+34 & erg s$^{-1}$   & -13.8627   & 2.8506E-06 & erg s$^{-1}$ cm$^{-2}$ \\
    \hline
    \multicolumn{7}{c}{$C_{\rm V} = C_{\rm Bol} -2.3826$ \qquad \qquad  $S_\lambda(V)$ from \cite{Landolt1992} \qquad \qquad $c_{\rm V} = c_{\rm Bol} -2.3826$} \\
    
    SI  & 68.8148 & 3.3568E+27 & W              & -21.3799   & 2.8506E-09 & W m$^{-2}$             \\
    cgs & 86.3147 & 3.3568E+34 & erg s$^{-1}$   & -13.8799   & 2.8506E-06 & erg s$^{-1}$ cm$^{-2}$ \\
    \bottomrule
    \end{tabular}
    \label{tab:8}
\end{table*}
%----------------------------------------------------------------------------------------------

\subsection{Interstellar extinctions from the sample spectra}

Interstellar extinction ($A_{\rm V}$) is a parameter like stellar $L$, which is not directly observable. Fortunately, both could be computed or estimated from the other observable parameters. Calculating $L$ of a star from its $T_{\rm eff}$ and $R$ is called the direct method by \cite{Eker2021b}. Primary aim of this study is to develop one of the indirect methods, which uses one of the bolometric corrections ($BC_{\rm B}$, $BC_{\rm V}$, $BC_{\rm R}$, ....) and show, this method permits one to obtain $L$ of a star from one of its apparent magnitudes ($B$, $V$, $R$, ...), extinctions ($A_{\rm B}$, $A_{\rm V}$, $A_{\rm R}$, ....) and its {\it Gaia} DR3 trigonometric parallax only; even without knowing its radius ($R$).

Here, in this study we argue that there is an alternative way to obtain $A_{\rm V}$ of a star directly from a single relation (Equation~(\ref{equ:19})) by imposing the definition of the extinction, $A_{\rm V}$, as the difference between the observed ($M_{\rm V}$(obs)) and the intrinsic ($M_{\rm V}$(int)) absolute visual magnitudes of the star, where the intrinsic absolute visual magnitude expressed as: $M_{\rm V}{\rm (int)} = M_{\rm Bol}-BC_{\rm V}$ according to Equation~(\ref{equ:1}). It appears simple and directly applicable, or as a shortcut to eliminate any other methods providing $A_{\rm V}$, including the SED analysis used in this study. Though a serious problem with this equation is that it requires $M_{\rm Bol}$ to be known in addition to $V$, $\varpi$, and $BC_{\rm V}$. The main purpose, however, is to obtain $M_{\rm Bol}$ by adding the missing part of the total radiation ($BC_{\rm V}$) to the intrinsic absolute visual magnitude ($M_{\rm V}$(int)). As if there is only a single equation with two unknowns to be solved. The good news is this: $M_{\rm Bol}$ of a star could be computed via Equation~(\ref{equ:3}) by an application of Stefan-Boltzmann law: $L=4\pi R^2 \sigma {T_{\rm eff}}^4$ that requires $R$ in addition to $T_{\rm eff}$ for a star. The bad news is this: an additional unknown ($R$) is introduced. It appears that there is no way to eliminate $R$ as an unknown unless a method is developed to obtain both $M_{\rm Bol}$ and $A_{\rm V}$ at last to confirm us $M_{\rm Bol} = M_{\rm V}+BC_{\rm V}$, where both of the absolute magnitudes are intrinsic. 

If $M_{\rm Bol}$ is calculated via Equation~(\ref{equ:3}) using $T_{\rm eff}$ and $R$ of the star, another problem arises because of observational uncertainties of the $T_{\rm eff}$ and $R$. Second and fourth powers associated with $R$ and $T_{\rm eff}$ in the Stefan-Boltzmann cause observational errors to propagate to enormous intolerable values as displayed in Figure~\ref{fig:t_r_l_uncertainty}. Such inflated uncertainties in $L$, consequently in $M_{\rm Bol}$, then most often one comes across a negative value for the parameter $A_{\rm V}$. This may be the main reason why many researchers prefer other methods rather than using Equation~(\ref{equ:19}) to obtain $A_{\rm V}$. 

\begin{figure*}
    \centering
    \includegraphics[width=0.45\linewidth]{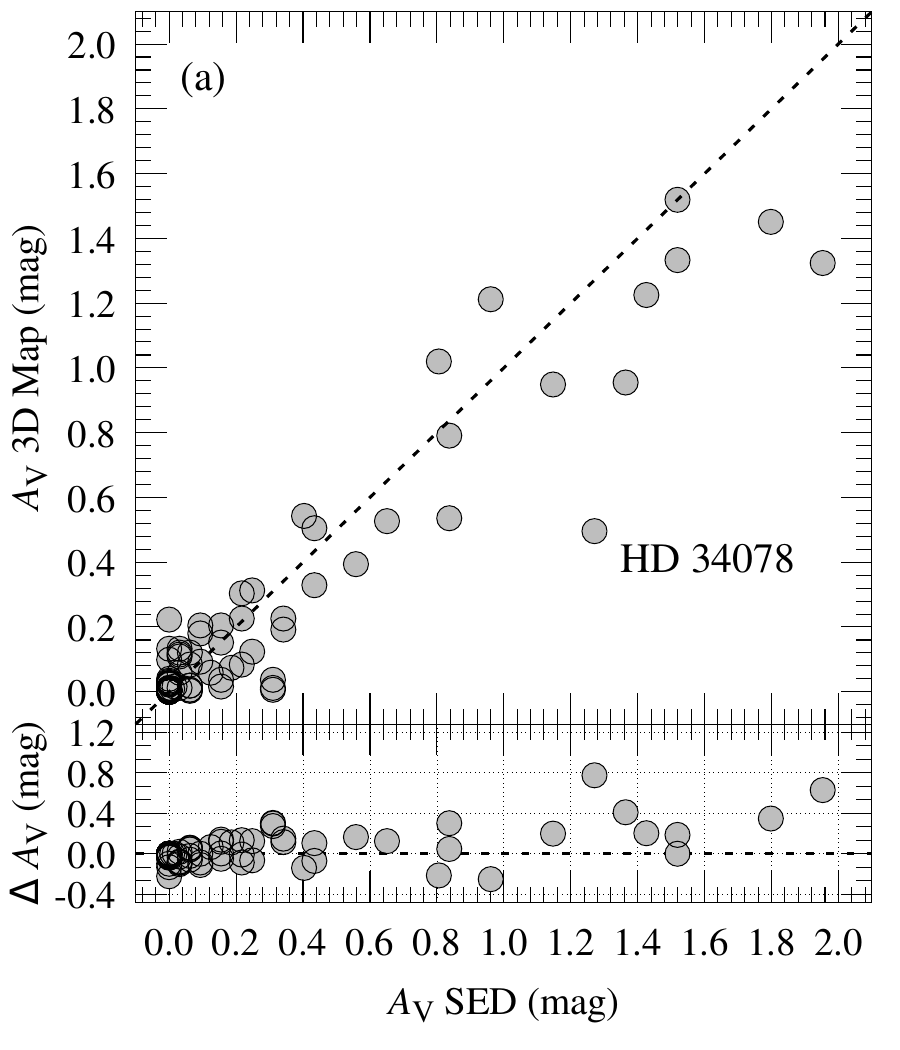}
    \includegraphics[width=0.45\linewidth]{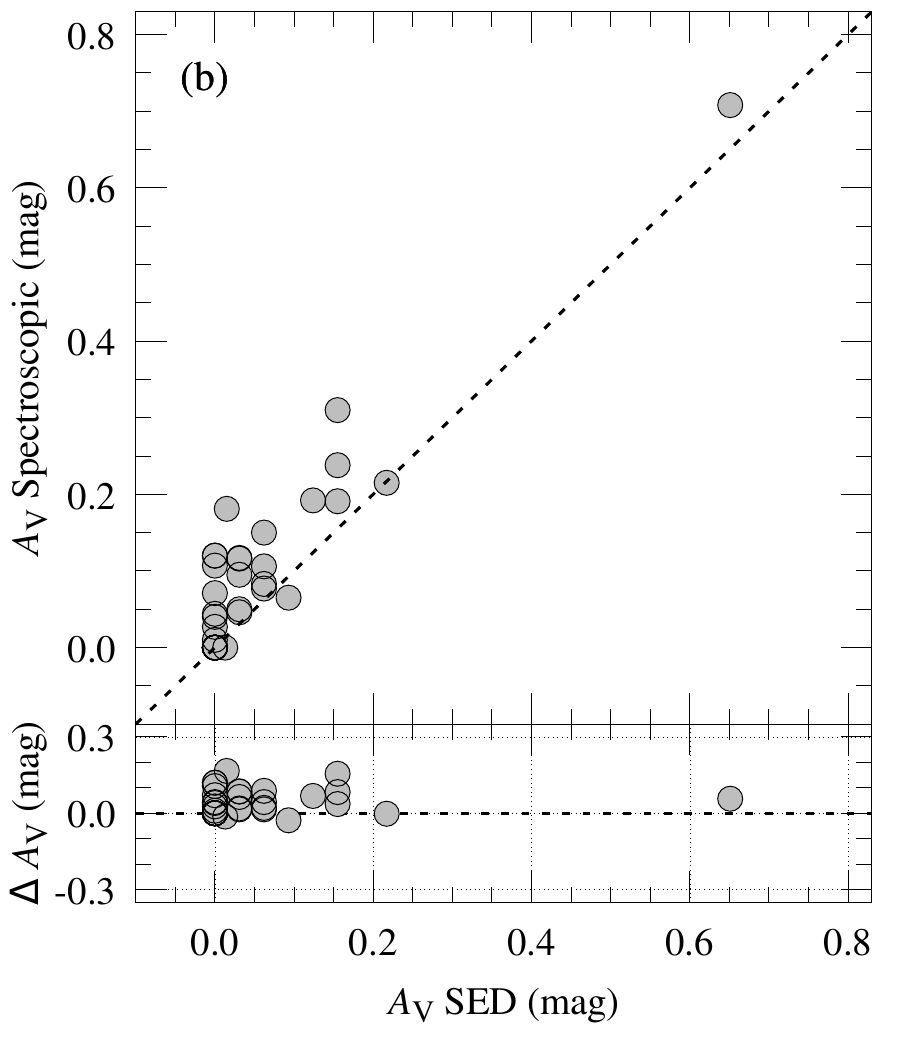}
    \caption{(a) Comparing interstellar extinctions ($A_{\rm V}$) of the SED method to the ones from the 3D Galactic dust maps \citep{Green2019}. (b) comparing computed spectroscopic $A_{\rm V}$ to $A_{\rm V}$ from the SED analysis.}
    \label{fig:A_v_comparison}
\end{figure*}

To eliminate erroneous $M_{\rm Bol}$, we have chosen the most accurate stars in our sample ($ \Delta L_{\rm V}/L < 11\%, \Delta \varpi / \varpi < 5\% \text{, and}\,\Delta V < 0.014$ mag) and applied Equation~(\ref{equ:19}) to them by using their spectroscopic $BC_{\rm V}$. The results are listed in Table~\ref{tab:9}. Because the negative $A_{\rm V}$ is not possible, all negative values are replaced by zero in column 16. Columns of Table~\ref{tab:9} are self-explanatory, indicating observational and propagated errors of the observed and computed quantities. Calculated $A_{\rm V}$ values are compared to the $A_{\rm V}$ values from the SED analysis and 3D Galactic maps\footnote{http://argonaut.skymaps.info/query} that are shown in Figure~\ref{fig:A_v_comparison}. 

%----------------------------------------------------------------------------------------------
%Tablo 9
\newpage
{\setlength{\tabcolsep}{2pt}
\renewcommand*{\arraystretch}{0.78}
\begin{longtable*}{llccccccccccccccc}
    \caption{Interstellar extinctions ($A_{\rm V}$) obtained directly from most accurate apparent magnitudes ($V$), trigonometric parallaxes ($\varpi$), and absolute bolometric magnitudes ($M_{\rm Bol}$), which comes from the effective temperature ($T_{\rm eff}$) and radii ($R$) by using spectroscopic ($BC_{\rm V}$).}\\
    \hline
    Order & Star & $V$ & err & $\varpi$ & err & err & $M_{\rm V}$(obs) & err & $M_{\rm Bol}$ & err & $BC_{\rm V}$ & err & $M_{\rm V}$(int) & err & $A_{\rm V}$ & err \\ 
         & & \multicolumn{2}{c}{(mag)} & (mas) & (\%) & (mag) & \multicolumn{2}{c}{(mag)} & \multicolumn{2}{c}{(mag)} & \multicolumn{2}{c}{(mag)} &\multicolumn{2}{c}{(mag)} & \multicolumn{2}{c}{(mag)} \\ \hline
        \endfirsthead
    \multicolumn{17}{c}%
{{\bfseries \tablename\ \thetable{}} -- continued from previous page} \\ 
\hline  
Order & Star & $V$ & err & $\varpi$ & err & err & $M_{\rm V}$(obs) & err & $M_{\rm Bol}$ & err & $BC_{\rm V}$ & err & $M_{\rm V}$(int) & err & $A_{\rm V}$ & err \\ 
         & & \multicolumn{2}{c}{(mag)} & (mas) & (\%) & (mag) & \multicolumn{2}{c}{(mag)} & \multicolumn{2}{c}{(mag)} & \multicolumn{2}{c}{(mag)} &\multicolumn{2}{c}{(mag)} & \multicolumn{2}{c}{(mag)} \\ \hline
\endhead
\hline \multicolumn{17}{c}{{Continued on next page}} \\ \hline
\endfoot
\hline 
\endlastfoot
        1 & HD\,172167 & 0.030 & 0.010 & 130.230 & 0.28 & 0.006 & 0.604 & 0.012 & 0.564 & 0.071 & -0.176 & 0.007 & 0.740 & 0.071 & 0 & 0.072 \\ 
        2 & HD\,62509 & 1.140 & 0.010 & 96.540 & 0.28 & 0.006 & 1.064 & 0.012 & 0.836 & 0.030 & -0.345 & 0.007 & 1.181 & 0.031 & 0 & 0.033 \\ 
        3 & HD\,102870 & 3.600 & 0.010 & 90.895 & 0.21 & 0.005 & 3.393 & 0.011 & 3.405 & 0.078 & -0.040 & 0.008 & 3.445 & 0.078 & 0 & 0.079 \\ 
        4 & HD\,124897 & -0.050 & 0.010 & 88.830 & 0.61 & 0.013 & -0.307 & 0.017 & -0.831 & 0.092 & -0.580 & 0.007 & -0.251 & 0.092 & 0 & 0.094 \\ 
        5 & HD\,146233 & 5.500 & 0.010 & 70.737 & 0.09 & 0.002 & 4.748 & 0.010 & 4.793 & 0.114 & -0.075 & 0.007 & 4.868 & 0.114 & 0 & 0.115 \\ 
        6 & HD\,117176 & 4.970 & 0.009 & 55.251 & 0.14 & 0.003 & 3.682 & 0.010 & 3.569 & 0.092 & -0.123 & 0.007 & 3.692 & 0.092 & 0 & 0.093 \\ 
        7 & HD\,113226 & 2.790 & 0.010 & 30.211 & 0.63 & 0.014 & 0.191 & 0.017 & -0.091 & 0.096 & -0.290 & 0.008 & 0.199 & 0.096 & 0 & 0.098 \\ 
        8 & HD\,32115 & 6.320 & 0.010 & 20.376 & 0.15 & 0.003 & 2.866 & 0.011 & 2.881 & 0.117 & 0.002 & 0.016 & 2.879 & 0.118 & 0 & 0.119 \\ 
        9 & HD\,141714 & 4.630 & 0.010 & 19.497 & 0.49 & 0.011 & 1.080 & 0.015 & 0.837 & 0.106 & -0.160 & 0.007 & 0.997 & 0.106 & 0.083 & 0.107 \\ 
        10 & HD\,45638 & 6.590 & 0.009 & 17.384 & 1.30 & 0.028 & 2.791 & 0.030 & 2.829 & 0.084 & 0.012 & 0.010 & 2.817 & 0.085 & 0 & 0.090 \\ 
        11 & HD\,162570 & 6.130 & 0.009 & 10.624 & 0.20 & 0.004 & 1.261 & 0.010 & 1.254 & 0.103 & 0.020 & 0.011 & 1.234 & 0.104 & 0.027 & 0.104 \\ 
        12 & HD\,18543 & 5.230 & 0.010 & 8.943 & 2.09 & 0.045 & -0.013 & 0.047 & -0.266 & 0.093 & -0.133 & 0.013 & -0.133 & 0.094 & 0.120 & 0.105 \\ 
        13 & HD\,37077 & 5.234 & 0.009 & 8.766 & 0.90 & 0.020 & -0.052 & 0.022 & 0.024 & 0.075 & 0.012 & 0.009 & 0.012 & 0.076 & 0 & 0.079 \\ 
        14 & HD\,99922 & 5.813 & 0.009 & 8.645 & 0.46 & 0.010 & 0.497 & 0.013 & 0.417 & 0.089 & -0.070 & 0.012 & 0.487 & 0.090 & 0.010 & 0.091 \\ 
        15 & HD\,18633 & 5.550 & 0.010 & 8.609 & 1.23 & 0.027 & 0.225 & 0.029 & -0.182 & 0.113 & -0.287 & 0.009 & 0.105 & 0.113 & 0.120 & 0.117 \\ 
        16 & HD\,150117 & 5.390 & 0.010 & 7.889 & 1.28 & 0.028 & -0.125 & 0.030 & -0.599 & 0.119 & -0.293 & 0.013 & -0.306 & 0.120 & 0.181 & 0.123 \\ 
        17 & HD\,52100 & 6.546 & 0.010 & 7.784 & 1.68 & 0.036 & 1.002 & 0.038 & 1.026 & 0.100 & 0.020 & 0.013 & 1.006 & 0.101 & 0 & 0.108 \\ 
        18 & HD\,29335 & 5.315 & 0.009 & 7.030 & 4.69 & 0.102 & -0.450 & 0.102 & -1.362 & 0.081 & -0.721 & 0.009 & -0.641 & 0.081 & 0.191 & 0.131 \\ 
        19 & HD\,1439 & 5.875 & 0.009 & 6.648 & 1.23 & 0.027 & -0.012 & 0.028 & -0.188 & 0.114 & -0.180 & 0.011 & -0.008 & 0.115 & 0 & 0.118 \\ 
        20 & HD\,35693 & 6.182 & 0.010 & 6.605 & 0.71 & 0.015 & 0.281 & 0.018 & 0.128 & 0.110 & -0.113 & 0.010 & 0.241 & 0.110 & 0.040 & 0.112 \\ 
        21 & HD\,32040 & 6.630 & 0.009 & 5.878 & 1.66 & 0.036 & 0.476 & 0.037 & 0.099 & 0.118 & -0.441 & 0.011 & 0.540 & 0.119 & 0 & 0.124 \\ 
        22 & HD\,78556 & 5.609 & 0.012 & 4.872 & 2.58 & 0.056 & -0.953 & 0.057 & -1.355 & 0.047 & -0.285 & 0.009 & -1.070 & 0.048 & 0.117 & 0.075 \\ 
        23 & HD\,63975 & 5.160 & 0.010 & 4.002 & 2.25 & 0.049 & -1.828 & 0.050 & -2.403 & 0.108 & -0.504 & 0.009 & -1.899 & 0.108 & 0.071 & 0.119 \\ 
        24 & HD\,27563 & 5.838 & 0.009 & 3.776 & 1.62 & 0.035 & -1.277 & 0.036 & -1.883 & 0.110 & -0.692 & 0.011 & -1.191 & 0.111 & 0 & 0.116 \\ 
        25 & HD\,40967 & 5.010 & 0.010 & 3.487 & 2.91 & 0.063 & -2.278 & 0.064 & -3.654 & 0.115 & -1.066 & 0.013 & -2.588 & 0.116 & 0.310 & 0.132 \\ 
        26 & HD\,122563 & 6.190 & 0.010 & 3.099 & 1.07 & 0.023 & -1.354 & 0.025 & -1.820 & 0.103 & -0.274 & 0.007 & -1.546 & 0.103 & 0.192 & 0.106 \\ 
        27 & HD\,174959 & 6.082 & 0.009 & 2.978 & 1.54 & 0.033 & -1.549 & 0.035 & -2.520 & 0.110 & -0.855 & 0.013 & -1.665 & 0.111 & 0.116 & 0.116 \\ 
        28 & HD\,46189 & 5.903 & 0.009 & 2.972 & 2.23 & 0.049 & -1.732 & 0.049 & -3.002 & 0.103 & -1.163 & 0.010 & -1.839 & 0.103 & 0.107 & 0.115 \\ 
        29 & HD\,1279 & 5.764 & 0.014 & 2.849 & 1.81 & 0.039 & -1.963 & 0.042 & -2.676 & 0.116 & -0.667 & 0.013 & -2.009 & 0.117 & 0.046 & 0.124 \\ 
        30 & HD\,36285 & 6.313 & 0.010 & 2.825 & 2.29 & 0.050 & -1.432 & 0.051 & -3.378 & 0.064 & -1.731 & 0.010 & -1.647 & 0.065 & 0.215 & 0.082 \\ 
        31 & HD\,45321 & 6.133 & 0.010 & 2.814 & 2.32 & 0.050 & -1.620 & 0.051 & -2.742 & 0.094 & -1.072 & 0.010 & -1.670 & 0.095 & 0.050 & 0.108 \\ 
        32 & HD\,35299 & 5.700 & 0.009 & 2.773 & 3.30 & 0.072 & -2.085 & 0.072 & -4.149 & 0.066 & -1.987 & 0.010 & -2.162 & 0.067 & 0.077 & 0.098 \\ 
        33 & HD\,36960 & 4.720 & 0.010 & 2.617 & 4.60 & 0.100 & -3.191 & 0.100 & -5.765 & 0.093 & -2.479 & 0.009 & -3.286 & 0.093 & 0.095 & 0.137 \\ 
        34 & HD\,58599 & 6.375 & 0.010 & 2.572 & 3.92 & 0.850 & -1.574 & 0.086 & -2.269 & 0.110 & -0.651 & 0.011 & -1.618 & 0.111 & 0.044 & 0.140 \\ 
        35 & HD\,37744 & 6.220 & 0.010 & 2.520 & 2.67 & 0.058 & -1.773 & 0.059 & -3.995 & 0.098 & -1.984 & 0.010 & -2.011 & 0.099 & 0.238 & 0.115 \\ 
        36 & HD\,36430 & 6.208 & 0.010 & 2.517 & 2.78 & 0.060 & -1.787 & 0.061 & -3.386 & 0.064 & -1.449 & 0.009 & -1.937 & 0.065 & 0.150 & 0.089 \\ 
        37 & HD\,32612 & 6.406 & 0.010 & 2.462 & 2.22 & 0.048 & -1.638 & 0.049 & -3.229 & 0.076 & -1.485 & 0.008 & -1.744 & 0.076 & 0.106 & 0.091 \\ 
        38 & HD\,37356 & 6.180 & 0.010 & 2.197 & 2.53 & 0.055 & -2.111 & 0.056 & -4.625 & 0.097 & -1.806 & 0.015 & -2.819 & 0.098 & 0.708 & 0.113 \\ 
        39 & HD\,55856 & 6.270 & 0.010 & 1.389 & 3.87 & 0.084 & -3.016 & 0.085 & -4.804 & 0.119 & -1.723 & 0.013 & -3.081 & 0.120 & 0.065 & 0.147 \label{tab:9}
\end{longtable*}}

%-----------------------------------------------------------------------------------------------
%Tablo 10

{\setlength{\tabcolsep}{1pt}
\renewcommand*{\arraystretch}{0.9}
\begin{table*}[ht]
\centering
\caption{Comparison of spectroscopic $A_{\rm V}$ with $A_{\rm V}$ (SED) and $A_{\rm V}$ (3D). The sequence of stars is like Table 9, from nearest to farthest.} 
\begin{tabular}{llcccclcccclccc}
\toprule
    ID & Star & $A_{\rm V}$ (Sp.) & $A_{\rm V}$ (SED) & $A_{\rm V}$ (3D) & ID & Star & $A_{\rm V}$ (Sp.) & $A_{\rm V}$ (SED) & $A_{\rm V}$ (3D) & ID & Star & $A_{\rm V}$ (Sp.) & $A_{\rm V}$ (SED) & $A_{\rm V}$ (3D)\\
    &   & \multicolumn{3}{c}{(mag)} &  & & \multicolumn{3}{c}{(mag)} & &  & \multicolumn{3}{c}{(mag)} \\
    \hline
        1 & HD\,172167 & 0 & 0 & 0.001         & 14 & HD\,99922 & 0.010 & 0 & 0.007      & 27 & HD\,174959 & 0.116 & 0.031 & 0.108 \\ 
        2 & HD\,62509 & 0 & 0 & 0.001          & 15 & HD\,18633 & 0.120 & 0 & 0.007      & 28 & HD\,46189 & 0.107 & 0 & 0.039 \\ 
        3 & HD\,102870 & 0 & 0 & 0.001         & 16 & HD\,150117 & 0.181 & 0.015 & 0.015 & 29 & HD\,1279 & 0.046 & 0.031 & 0.133 \\ 
        4 & HD\,124897 & 0 & 0 & 0.001         & 17 & HD\,52100 & 0 & 0 & 0.009          & 30 & HD\,36285 & 0.215 & 0.217 & 0.084 \\ 
        5 & HD\,146233 & 0 & 0 & 0             & 18 & HD\,29335 & 0.191 & 0.155 & 0.016  & 31 & HD\,45321 & 0.050 & 0.031 & 0.060 \\ 
        6 & HD\,117176 & 0 & 0 & 0.002         & 19 & HD\,1439 & 0.000 & 0.013 & 0.013   & 32 & HD\,35299 & 0.077 & 0.062 & 0.085 \\ 
        7 & HD\,113226 & 0 & 0 & 0.003         & 20 & HD\,35693 & 0.040 & 0 & 0.033      & 33 & HD\,36960 & 0.095 & 0.031 & 0.122 \\ 
        8 & HD\,32115 & 0 & 0 & 0.004          & 21 & HD\,32040 & 0 & 0 & 0.032          & 34 & HD\,58599 & 0.044 & 0 & 0.024 \\ 
        9 & HD\,141714 & 0.083 & 0.062 & 0.006 & 22 & HD\,78556 & 0.117 & 0.031 & 0.012  & 35 & HD\,37744 & 0.238 & 0.155 & 0.153 \\ 
        10 & HD\,45638 & 0 & 0 & 0.004         & 23 & HD\,63975 & 0.071 & 0 & 0.013      & 36 & HD\,36430 & 0.150 & 0.062 & 0.018 \\ 
        11 & HD\,162570 & 0.027 & 0 & 0.025    & 24 & HD\,27563 & 0 & 0 & 0.028          & 37 & HD\,32612 & 0.106 & 0.062 & 0.122 \\ 
        12 & HD\,18543 & 0.120 & 0 & 0.007     & 25 & HD\,40967 & 0.310 & 0.155 & 0.036  & 38 & HD\,37356 & 0.708 & 0.651 & 0.526 \\ 
        13 & HD\,37077 & 0 & 0 & 0.007         & 26 & HD\,122563 & 0.192 & 0.124 & 0.058 & 39 & HD\,55856 & 0.065 & 0.093 & 0.093 \\

\bottomrule
\end{tabular}
\label{tab:10}

\end{table*}}
%----------------------------------------------------------------------------------------------

%----------------------------------------------------------------------------------------------

Having very accurate spectroscopic $BC_{\rm V}$, the method of using analytical relation (Equation~(\ref{equ:19})) appears successful according to the data in Table~\ref{tab:10}, where the negative $A_{\rm V}$ is not many, and according to Figure~\ref{fig:A_v_comparison}, where $A_{\rm V}$ from equation and $A_{\rm V}$ from the SED analysis and $A_{\rm V}$ from the 3D maps were compared. Therefore, the computed $A_{\rm V}$ values in Table~\ref{tab:10} and Figure~\ref{fig:A_v_comparison} confirm that it is useful to compute interstellar extinctions if observational data, including $T_{\rm eff}$ and $R$, are sufficiently accurate and precise otherwise unusable because of the negative values which are inevitable due to intolerable uncertainties associated with the observational parameters. 

\subsection{Limiting and typical accuracies of $M_{\rm Bol}$ and $L$ by spectroscopic $BC$}

If a star is in the Local Bubble \citep{Leeroy1993, Lallement2019}, where interstellar extinction could be ignored, for the indirect method of obtaining $L$ using a spectroscopic $BC_{\rm V}$ as described in this study, there could only be one dominant source of uncertainty in the first step for calculating $M_{\rm Bol}$ from $M_{\rm V} + BC_{\rm V}$ where both quantities could be accurate within milimagnitudes in today's technology, and this dominant source most probably is the star's trigonometric parallax $\varpi$.

It is clear that the $5\log_{10}{e}$ factor in front of the relative error of  $\varpi$, which transfers it to magnitudes, inflates the uncertainty contribution of $\varpi$, while such an inflating factor does not exist for the apparent magnitude ($\Delta m_{\rm V}$) and the extinction ($\Delta A_{\rm V}$) according to Equation~(\ref{equ:10}). It is also known that a relative uncertainty in $\varpi$ doubles in the propagation process, so that the uncertainty in $L$ would be twice the uncertainty of $\varpi$. This is because the stellar flux measured just above the Earth’s atmosphere is inversely proportional to the squares of the star's distance.

Therefore, assuming that the contributions of $\Delta m_{\rm V}$ and $\Delta A_{\rm V}$ are negligible compared to the uncertainty in the distance (or $\varpi$)—taken here as 4\%, 6\%, or 10\% as illustrative examples—and that the uncertainty in the spectroscopic $BC_{\rm V}$ is negligible (see Table~\ref{tab:11}), the corresponding uncertainties in the predicted $L$ are 8\%, 12\%, and 20\%, respectively. These are equivalent to errors of $\pm$0.0869, $\pm$0.1303, and $\pm$0.2171 mag in the magnitude scale, according to Equation~(\ref{equ:3}) for the absolute bolometric magnitude of the same star.

Histogram distribution of the parallax errors (not displayed) for the stars in the present sample (Table~\ref{tab:3}) indicates that the mean value of the $\varpi$ errors is 2.16\% with a standard deviation of 2.58\%. Taking this as a typical uncertainty in the parallax, the typical error of the predicted $L$ would be 5.16\%. This is a clear-cut improvement compared to the typical uncertainties of $L$ announced previously by \citet{Eker2021b}, which is 8.2\% – 12.2\%. Using a standard $BC_{\rm V}$ removes extra 10\% uncertainty of non-standard $BC$ \citep{Torres2010, Eker2021b}, and then using spectroscopic $BC_{\rm V}$ rather than standard photometric $BC_{\rm V}$, obviously made this difference of further improvement in the accuracy of the predicted $L$ for the stars in the Local Bubble.

Unfortunately, starlight is subject to interstellar extinction ($A_{\rm V}$). The smallest $A_{\rm V}$ error in Table~\ref{tab:9} belongs to the star HD 62509, which has $\pm$0.033 mag uncertainty. Therefore, the next biggest contribution to the uncertainty of a predicted $L$ comes from the parameter $A_{\rm V}$. It may even dominate $\varpi$ uncertainty as in the star HD 62509 with 0.28\%, which is $\pm$0.006 mag in the magnitude scale. Spectroscopic $BC_{\rm V}$ of this star has a $\pm$0.001 mag uncertainty in Table~\ref{tab:11}. Very bright stars such as HD 62509 ($\beta$ Gem), and HD 124897 ($\alpha$ Boo, Arcturus) are recorded in the SIMBAD database without an error in their visual brightness; thus, the $V$ errors of such stars were taken $\pm$0.01 mag in this study. Consequently, $\pm$0.033 mag in $A_{\rm V}$, $\pm$0.01 mag in $V$, $\pm$0.006 in $\varpi$, and $\pm$0.001 mag in the spectroscopic $BC_{\rm V}$, implies that the error in the $M_{\rm Bol}$ of HD 62509 is $\pm$0.0350 mag, which corresponds to 3.2\% uncertainty in the $L$, which is the highest (limiting) accuracy among the sample stars in this study. 

The accuracy of $A_{\rm V}$ is usually worse than the accuracy of visual apparent magnitudes because its definition involves four brightness measurements, where the two of them form its uncorrected color and the other two form its intrinsic color. Therefore, the limiting accuracy of $L$, definitely, will depend mostly on the accuracy of interstellar extinction if the other contributions are at the level of milimagnitudes. For example, provided with  $\pm$0.01 mag in $A_{\rm V}$, $\pm$0.005 mag in $V$, and $\pm$0.005 in both  $\varpi$ and $BC_{\rm V}$, one can calculate $L$ of a star to have a 1.2 \%  limiting accuracy only by a single channel photometry, that is using only spectroscopic $BC_{\rm V}$ produced in this study. Multi-band spectroscopic $BC$, if produced in the future, could be used to improve it even further to achieve a stellar $L$, which is even more accurate than 1\%.

%---------------------------------------------------------------------------------------------

\section{Conclusion} \label{sec:conclusion}

It has been shown that obtaining $L$ of a star as accurate as 1\% or better by the indirect method with bolometric correction is now possible in today’s technology if one uses spectroscopic $BC_{\rm V}$ from high-resolution high \textit{S/N} ratio spectra.

Table~\ref{tab:11} presents values of spectroscopic $BC_{\rm V}$  from 128 sample spectra, where uncertainties at milimagnitudes are common. This precision as shown in Table~\ref{tab:11} was possible with the help of the zero-point constant determined in this study for the $BC_{\rm V}$ scale: $C_{\rm 2}=2.3653\pm0.0067$ mag if the $S_\lambda(V)$ profile function of \cite{Bessell1990} was used, or $C_{\rm 2}=2.3826\pm0.0076$ mag if the $S_\lambda(V)$ profile function of \cite{Landolt1992} was used.

An equation like $M_{\rm V}=2.5\log{L_{\rm V}}+C_{\rm V}$ and/or an equation like $V=2.5\times\log{f_{\rm V}}+c_{\rm V}$, where there are two unknowns in a single equation, could not easily be solved before this study. Now, both equations are solvable for $L_{\rm V}$ and/or $f_{\rm V}$ even for a single star if its apparent and absolute visual magnitude is known with the help of the zero-point constants $C_{\rm V}$ and $c_{\rm V}$ given in Table~\ref{tab:8}.  This success definitely comes from the value of the zero-point constants: $C_{\rm Bol} = 71.197425 ...$ mag if $L$ is in SI units for the absolute bolometric magnitudes ($M_{\rm Bol}$) and $c_{\rm Bol} = -18.997351 ...$ mag if the irradiance $f$ is in SI units, which were fixed by the General Assembly Resolution of IAU in 2015 (IAU2015GARB2).

It is possible that a high-resolution high \textit{S/N} ratio spectrum of a star is not available or technically not possible despite its $BC_{\rm V}$ is needed for calculating its $L$, $L_{\rm V}$ or visual to bolometric luminosity ratio ($L_{\rm V}/L$) from its apparent visual brightness ($V$), trigonometric parallax ($\varpi$) and interstellar extinction ($A_{\rm V}$). What best could be done for such cases is to use one of the two spectroscopic $BC_{\rm V}-T_{\rm eff}$ relations in Table~\ref{tab:7} that could provide a spectroscopic $BC_{\rm V}$ from its $T_{\rm eff}$. The third $BC_{\rm V}-T_{\rm eff}$ relation in Table~\ref{tab:7} obtained from $M_{\rm Bol}$, $M_{\rm V}$ and $T_{\rm eff}$ of the present sample is just a side product and/or a tool to see the difference between photometrically and spectroscopically determined $BC_{\rm V}-T_{\rm eff}$ relations.

An analytical formula using $M_{\rm Bol}$ and $BC_{\rm V}$ for computing $A_{\rm V}$, interstellar extinction of a star, from its unreddened apparent visual magnitude ($V$) and trigonometric parallax ($\varpi$) was tested. A limited number (39) of analytically computed $A_{\rm V}$ is compared to $A_{\rm V}$ estimated from the SED analysis of this study and 3D maps of \citet{Lallement2019}. We can conclude from these comparisons that the analytical formulae suggested in this study could safely be used for stars with accurately known $M_{\rm Bol}$. Otherwise, the large observational errors would dilute the computed $A_{\rm V}$ with meaningless negative values.

Having 18\% of the sample as giants and sub-giants, this study shows that the computed spectroscopic $BC_{\rm V}$ values do not indicate differences between luminosity classes or the $ \log{g}$ effect. Thus, for future work, we encourage investigations about how the metal abundance of a star and/or the low resolution of the spectrum would affect the values computed $BC_{\rm V}$. We also encourage interested researchers to obtain spectroscopic relations $BC$ and $BC-T_{\rm eff}$ for the other bands such as Johnson, $B$, $V$, $R$, $I$, ... and {\it Gaia} $G$, $G_{\rm BP}$, $G_{\rm RP}$ and other commonly used ones in the literature. 

\renewcommand*{\arraystretch}{0.85}
\begin{table*}[ht]
\centering
\footnotesize
\caption{Spectroscopic $BC_{\rm V}$ from the current sample of 128 stars and their errors for each $S_\lambda(V)$ profile.} 
\begin{tabular}{llcccllccc}
\toprule
    ID & Star & $BC_{\rm V}$\,$^b$ & $BC_{\rm V}$\,$^l$ & err & ID & Star & $BC_{\rm V}$\,$^b$ & $BC_{\rm V}$\,$^l$ & err\\
    &   & (mag) & (mag) &(mag) & & & (mag) & (mag) & (mag) \\  
    \midrule
         1 &    Sun        & -0.082 & -0.097 & 0.001 & 65  &    HD84937     &  0.043 & 0.039 & 0.008 \\ 
         2 &    HD1279     & -0.667 & -0.650 & 0.011 & 66  &    HD85503     & -0.556 & -0.594 & 0.002 \\ 
         3 &    HD1404     & -0.092 & -0.085 & 0.007 & 67  &    HD89021     & -0.145 & -0.137 & 0.005 \\ 
         4 &    HD1439     & -0.180 & -0.170 & 0.008 & 68  &    HD90882     & -0.304 & -0.292 & 0.008 \\ 
         5 &    HD2729     & -0.776 & -0.758 & 0.006 & 69  &    HD99922     & -0.070 & -0.063 & 0.010 \\ 
         6 &    HD4778     & -0.233 & -0.224 & 0.003 & 70  &    HD101364    & -0.073 & -0.088 & 0.005 \\ 
         7 &    HD10362    & -0.808 & -0.790 & 0.008 & 71  &    HD102870    & -0.040 & -0.052 & 0.005 \\ 
         8 &    HD10700    & -0.126 & -0.145 & 0.001 & 72  &    HD103095    & -0.134 & -0.153 & 0.003 \\ 
         9 &    HD12929    & -0.417 & -0.450 & 0.001 & 73  &    HD107328    & -0.434 & -0.468 & 0.002 \\ 
        10 &    HD15318    & -0.347 & -0.334 & 0.003 & 74  &    HD113226    & -0.290 & -0.317 & 0.003 \\ 
        11 &    HD16440    & -1.414 & -1.393 & 0.012 & 75  &    HD117176    & -0.123 & -0.141 & 0.002 \\ 
        12 &    HD18543    & -0.133 & -0.125 & 0.011 & 76  &    HD120198    & -0.281 & -0.270 & 0.003 \\ 
        13 &    HD18633    & -0.287 & -0.275 & 0.005 & 77  &    HD120315    & -1.161 & -1.141 & 0.009 \\ 
        14 &    HD18884    & -1.086 & -1.142 & 0.001 & 78  &    HD122563    & -0.274 & -0.298 & 0.003 \\ 
        15 &    HD19736    & -0.861 & -0.842 & 0.010 & 79  &    HD124897    & -0.580 & -0.620 & 0.001 \\ 
        16 &    HD22049    & -0.255 & -0.281 & 0.001 & 80  &    HD125924    & -1.650 & -1.628 & 0.015 \\ 
        17 &    HD22879    & -0.010 & -0.020 & 0.004 & 81  &    HD128167    &  0.021 &  0.016 & 0.003 \\ 
        18 &    HD23300    & -0.792 & -0.774 & 0.008 & 82  &    HD128311    & -0.313 & -0.344 & 0.006 \\ 
        19 &    HD27295    & -0.431 & -0.417 & 0.006 & 83  &    HD131873    & -0.792 & -0.840 & 0.002 \\ 
        20 &    HD27563    & -0.692 & -0.675 & 0.009 & 84  &    HD133208    & -0.228 & -0.252 & 0.001 \\ 
        21 &    HD27778    & -0.945 & -0.927 & 0.006 & 85  &    HD135742    & -0.646 & -0.629 & 0.007 \\ 
        22 &    HD29138    & -1.992 & -1.969 & 0.009 & 86  &    HD140283    & -0.002 & -0.011 & 0.004 \\ 
        23 &    HD29139    & -0.925 & -0.979 & 0.001 & 87  &    HD141714    & -0.160 & -0.179 & 0.002 \\ 
        24 &    HD29335    & -0.721 & -0.704 & 0.006 & 88  &    HD146233    & -0.075 & -0.090 & 0.002 \\ 
        25 &    HD29589    & -0.818 & -0.800 & 0.005 & 89  &    HD148379    & -1.164 & -1.144 & 0.006 \\ 
        26 &    HD32040    & -0.441 & -0.427 & 0.009 & 90  &    HD148688    & -1.616 & -1.595 & 0.007 \\ 
        27 &    HD32115    &  0.002 &  0.000 & 0.014 & 91  &    HD150117    & -0.293 & -0.281 & 0.011 \\ 
        28 &    HD32309    & -0.280 & -0.268 & 0.009 & 92  &    HD160762    & -1.315 & -1.294 & 0.006 \\ 
        29 &    HD32612    & -1.485 & -1.410 & 0.005 & 93  &    HD162570    &  0.020 & 0.021  & 0.009 \\ 
        30 &    HD32630    & -1.282 & -1.261 & 0.010 & 94  &    HD171301    & -0.507 & -0.492 & 0.007 \\ 
        31 &    HD34078    & -2.663 & -2.638 & 0.007 & 95  &    HD171432    & -1.912 & -1.889 & 0.015 \\ 
        32 &    HD34310    & -0.331 & -0.318 & 0.011 & 96  &    HD172167    & -0.176 & -0.166 & 0.001 \\ 
        33 &    HD35299    & -1.987 & -1.964 & 0.007 & 97  &    HD174959    & -0.855 & -0.837 & 0.011 \\ 
        34 &    HD35497    & -0.768 & -0.751 & 0.001 & 98  &    HD177756    & -0.485 & -0.470 & 0.003 \\ 
        35 &    HD35693    & -0.113 & -0.105 & 0.007 & 99  &    HD186427    & -0.078 & -0.093 & 0.002 \\ 
        36 &    HD35912    & -1.513 & -1.492 & 0.007 & 100 &    HD186791    & -0.741 & -0.786 & 0.002 \\ 
        37 &    HD36285    & -1.731 & -1.709 & 0.007 & 101 &    HD189319    & -0.944 & -0.997 & 0.006 \\ 
        38 &    HD36430    & -1.449 & -1.428 & 0.006 & 102 &    HD189741    & -0.087 & -0.079 & 0.006 \\ 
        39 &    HD36512    & -2.868 & -2.843 & 0.008 & 103 &    HD189957    & -2.757 & -2.732 & 0.009 \\ 
        40 &    HD36591    & -2.292 & -2.268 & 0.008 & 104 &    HD192263    & -0.301 & -0.331 & 0.004 \\ 
        41 &    HD36960    & -2.479 & -2.454 & 0.007 & 105 &    HD193183    & -1.592 & -1.570 & 0.013 \\ 
        42 &    HD37077    &  0.012 &  0.011 & 0.006 & 106 &    HD195556    & -1.247 & -1.227 & 0.007 \\ 
        43 &    HD37356    & -1.806 & -1.783 & 0.014 & 107 &    HD196740    & -0.955 & -0.936 & 0.006 \\ 
        44 &    HD37744    & -1.984 & -1.961 & 0.007 & 108 &    HD197512    & -1.937 & -1.914 & 0.011 \\ 
        45 &    HD40967    & -1.066 & -1.047 & 0.011 & 109 &    HD201091    & -0.562 & -0.606 & 0.002 \\ 
        46 &    HD45321    & -1.072 & -1.053 & 0.008 & 110 &    HD201092    & -0.692 & -0.739 & 0.003 \\ 
        47 &    HD45638    &  0.012 &  0.010 & 0.007 & 111 &    HD205139    & -2.271 & -2.247 & 0.010 \\ 
        48 &    HD46189    & -1.163 & -1.143 & 0.007 & 112 &    HD207538    & -2.695 & -2.670 & 0.007 \\ 
        49 &    HD47100    & -0.673 & -0.657 & 0.008 & 113 &    HD208266    & -2.070 & -2.047 & 0.012 \\ 
        50 &    HD48843    & -0.054 & -0.050 & 0.004 & 114 &    HD209419    & -0.900 & -0.882 & 0.008 \\ 
        51 &    HD48915    & -0.209 & -0.199 & 0.004 & 115 &    HD209975    & -2.829 & -2.804 & 0.005 \\ 
        52 &    HD49933    &  0.022 &  0.017 & 0.002 & 116 &    HD212061    & -0.251 & -0.239 & 0.005 \\ 
        53 &    HD52100    &  0.020 &  0.021 & 0.011 & 117 &    HD213087    & -2.074 & -2.051 & 0.008 \\ 
        54 &    HD54764    & -1.412 & -1.391 & 0.006 & 118 &    HD214263    & -1.633 & -1.611 & 0.014 \\ 
        55 &    HD55856    & -1.723 & -1.700 & 0.011 & 119 &    HD214923    & -0.528 & -0.512 & 0.006 \\ 
        56 &    HD55879    & -2.553 & -2.528 & 0.010 & 120 &    HD214994    & -0.138 & -0.129 & 0.008 \\ 
        57 &    HD58599    & -0.651 & -0.635 & 0.008 & 121 &    HD215191    & -1.765 & -1.742 & 0.010 \\ 
        58 &    HD62509    & -0.345 & -0.374 & 0.001 & 122 &    HD217014    & -0.096 & -0.113 & 0.001 \\ 
        59 &    HD63975    & -0.504 & -0.489 & 0.005 & 123 &    HD218045    & -0.352 & -0.339 & 0.006 \\ 
        60 &    HD71155    & -0.185 & -0.175 & 0.007 & 124 &    HD220009    & -0.584 & -0.624 & 0.002 \\ 
        61 &    HD78556    & -0.285 & -0.273 & 0.006 & 125 &    HD220825    & -0.291 & -0.281 & 0.007 \\ 
        62 &    HD82106    & -0.356 & -0.390 & 0.006 & 126 &    HD222173    & -0.458 & -0.444 & 0.007 \\ 
        63 &    HD82621    & -0.098 & -0.090 & 0.006 & 127 &    HD222661    & -0.365 & -0.351 & 0.006 \\ 
        64 &    HD82943    & -0.066 & -0.079 & 0.006 & 128 &    HD222762    & -0.597 & -0.581 & 0.007 \\
        \bottomrule
        \multicolumn{10}{l}{(\textit{b}) \cite{Bessell1990} and (\textit{l}) \cite{Landolt1992}}
\end{tabular}
\label{tab:11}
\end{table*}

%---------------------------------------------------------------------------------------------

\section*{Acknowledgments}
We thank anonymous referees for their insightful and constructive suggestions, which significantly improved the paper. We thank T\"{U}B{\.{I}}TAK for funding this research under project number 123C161. Funding was provided by the Scientific Research Projects Coordination Unit of Istanbul University as project number 40044. This research has made use of NASA's Astrophysics Data System. The VizieR and Simbad databases at CDS, Strasbourg, France were invaluable for the project as were data from the European Space Agency (ESA) mission \emph{Gaia}\footnote{https://www.cosmos.esa.int/gaia}, processed by the \emph{Gaia} Data Processing and Analysis Consortium (DPAC)\footnote{https://www.cosmos.esa.int/web/gaia/dpac/consortium}. Funding for DPAC has been provided by national institutions, in particular, the institutions participating in the \emph{Gaia} Multilateral Agreement. 

\software{\texttt{Astropy} \citep{astro1,astro2,astro3}, \texttt{Matplotlib} \citep{matplotlib}, \texttt{NumPy} \citep{numpy}, \texttt{SciPy} \citep{scipy}.} 
\newpage
%---------------------------------------------------------------------------------------------
% \bibliography{sample631}{}
\bibliography{reference}
\bibliographystyle{aasjournal}

\appendix
\section{ A method of obtaining zero-point constants for visual magnitudes} \label{sec:Appendix}

Relative photometry has a great advantage in that its users do not need to know the zero-point constants to determine apparent and absolute magnitudes of stars, as long as brightness comparisons fall within the same wavelength range or in the same filter. Despite the opposite is not true, astronomers artificially assumed intrinsic colors ($U-B$, $B-V$, $V-R$, …) of Vega are zero as if each intrinsic color is another independent magnitude system in addition to single band magnitude systems such as $U$, $B$, $V$, $R$ etc., even though they appear as differences of stellar magnitudes at two different bands. If a nearby star has equal magnitudes at two filters, this does not mean it has the same brightness at these two filters, but the same effective temperature (10\,000 K) as Vega.

Absolute photometry, however, requires the zero-point constants to be known for obtaining actual luminosity differences at two different bands. Consider the question: What fraction of a stellar luminosity is emitted within the visual wavelengths? To be able to answer this question, Equation~(\ref{equ:3}) must be adopted for the visual band first as:

\begin{equation}
    M_{\rm V} = 2.5\log{L_{\rm V}+C_{\rm V}}
    \label{equ:4}
\end{equation}

\noindent where $M_{\rm V}$ is the absolute visual magnitude representing the visual part of the total luminosity symbolized by $L_{\rm V}$ and $C_{\rm V}$ is the zero-point constant for the absolute visual magnitudes. Because $C_{\rm V}$ cancels automatically if $M_{\rm V}$ of two stars are subtracted, there was no need for a unique value like $C_{\rm Bol} = 71.197425...$, if $L$ is in SI units, which was assigned to the zero-point constant of the absolute bolometric magnitudes by IAU2015GARB2. Furthermore, Equation~(\ref{equ:4}) is unsolvable for the two unknowns ($L_{\rm V}$ and $C_{\rm V}$) from a single $M_{\rm V}$ and $C_{\rm V}$ vanishes if two $M_{\rm V}$ are available from two stars. The zero-point constants defined for AB and ST magnitudes \citep{Casagrande2014, Bessell1998} are also useless because they are for monochromatic brightness. Fortunately, IAU2015GARB2 gave us an opportunity here in this study that we are now able to describe a method for determining zero-point constants empirically for the Vega system of magnitudes using the two equations below:

\begin{equation}
    BC_{\rm V}=M_{\rm Bol}-M_{\rm V}=2.5\log{\frac{L_{\rm V}}{L}}+\left(C_{\rm Bol}-C_{\rm V}\right)=2.5\log{\frac{\int_{0}^{\infty} S_\lambda(V)F_\lambda d\lambda}{\int_{0}^{\infty} F_\lambda d\lambda}}+C_{\rm 2}
    \label{equ:5}
\end{equation}

\begin{equation}
    BC_{\rm V}=m_{\rm Bol}-m_{\rm V}=2.5\log{\frac{f_{\rm V}}{f}}+\left(c_{\rm Bol}-c_{\rm V}\right)=2.5\log{\frac{\int_{0}^{\infty} S_\lambda(V)f_\lambda d\lambda}{\int_{0}^{\infty} f_\lambda d\lambda}}+C_{\rm 2}
    \label{equ:6}
\end{equation}

\noindent where the first one could be obtained by subtracting Equation~(\ref{equ:4}) from Equation~(\ref{equ:3}) and the next one is the expression of \textit{BC} for the same star by using its apparent magnitudes \citep{Kuiper1938, Eker2025} in which \(C_{\rm 2}\) is the zero-point constant for the \( BC_{\rm V} \) scale. The subscript two indicates that $C_2$ is not just a constant but a constant made up of the two zero-point constants, because definite integrals are always written without one. For example, \(f=\int_{0}^{\infty} f_\lambda d\lambda\) and \(f_{\rm V}=\int_{0}^{\infty}S_\lambda(V) f_\lambda d\lambda\) are the bolometric and the visual fluxes reaching Earth from the star, respectively, if there is no extinction. Consequently, \(F=\int_{0}^{\infty}F_\lambda d\lambda\) and \(F_{\rm V}=\int_{0}^{\infty}S_\lambda(V)F_\lambda d\lambda\) are the bolometric and visual fluxes on the hypothetical surface of the star that using blackbody approximation and the definition of the effective temperature allows us to write \(F=\sigma T_{\rm eff}^4\). \(S_\lambda (V)\) is the transition profile of the visual filter, allowing visual photons only. Equation~(\ref{equ:6}) indicates \(C_{\rm 2}=c_{\rm Bol}-c_{\rm V}\), in which \(c_{\rm V}\) is the zero-point constant for Vega system of apparent visual and \(c_{\rm Bol}\) is the zero-point constant of apparent bolometric magnitudes, which is also given IAU2015GARB2 as \(c_{\rm Bol} =  -18.997 351...\) mag. Assuming stars radiate isotropically, one can also deduce \(\frac{L_{\rm V}}{L}=\frac{f_{\rm V}}{f}\) and \(C_{\rm 2}=C_{\rm Bol}-C_{\rm V}=c_{\rm Bol}-c_{\rm V}\) from Equations (\ref{equ:5}) and (\ref{equ:6}). The following are the steps for calculating \(C_{\rm 2}\) and its uncertainty for a star:

{\bf Step 1)} Calculate \(BC_{\rm V}\) of a star according to Equation~(\ref{equ:5}), where \(M_{\rm Bol}\) must be calculated by Equation~(\ref{equ:3}) using the Stefan-Boltzmann law, \(L=4\pi R^2T^4\), and  \(M_{\rm V}\) is from 

\begin{equation}
    M_{\rm V}=V+5\log{\varpi}+5-A_{\rm V}
\label{equ:7}
\end{equation}

\noindent where \textit{V} is the apparent visual magnitude, \(\varpi\) is the trigonometric parallax in arc seconds, and \(A_{\rm V}\) is the extinction in the \textit{V} band, which could be ignored if the star is in the Local Bubble \citep{Leeroy1993, Lallement2019}. Otherwise, there could be various methods to estimate it using Galactic dust maps \citep[e.g.,][]{Bilir2008, Schlafly2011, Green2019} or SED analysis as described by \cite{Bakis2022}.  

The uncertainty of \(BC_{\rm V}\) could be estimated by propagating observational errors as follows: 

\begin{equation}
    \Delta BC_{\rm V} = \sqrt{(\Delta M_{\rm Bol})^2+(\Delta M_{\rm V})^2}
    \label{equ:8}
\end{equation}

\noindent in which,

\begin{equation}
    \Delta M_{\rm Bol} = 1.0857 \times\left(\sqrt{\left(2\frac{\Delta R}{R}\right)^2 + \left(4\frac{\Delta T_{\rm eff}}{T_{\rm eff}}\right)^2}\right)
    \label{equ:9}
\end{equation}

\begin{equation}
    \Delta M_{\rm V}=\sqrt{\left(\Delta m_{\rm V}\right)^2+\left(2.1715 \times\frac{\sigma_\varpi}{\varpi}\right)^2+\left(\Delta A_{\rm V}\right)^2}
    \label{equ:10}
\end{equation}

\noindent where observational uncertainties of \textit{R} and \(T_{\rm eff}\) contribute through \(\Delta M_{\rm Bol}\) while uncertainties from the apparent magnitude (\textit{V}), trigonometric parallax (\(\varpi\)) and interstellar extinction ($A_{\rm V}$) contribute through \(\Delta M_{\rm V}\). 

\textbf{Step 2)} After a standard procedure of obtaining a spectrum by a spectrograph attached to the telescope, here we suggest using a normalized spectrum to avoid spectral features due to interstellar extinction. Thus, normalization of the continuum to one is very important for the method. The wavelength profile of the visual filter \(S_\lambda (V)\) must also be normalized to one. The resolution and \textit{S/N} of the spectrum could be optional for private or special purposes, but the spectrum must cover the wavelength range of the visual filter.  

Calculating the fractional luminosity \((L_{\rm V}/L)\) of the star from its spectrum becomes possible after multiplying the normalized spectrum by the Planck function of the effective temperature of the star, 
\begin{equation}
    B_\lambda(T_{\rm eff})=\frac{2hc^2}{\lambda^5}\left(e^{hc/\lambda kT_{\rm eff}}-1\right)^{-1}.
    \label{equ:11}
\end{equation}

Multiplying the normalized spectrum by the Planck function is not for restoring the observed spectrum, but for preparing the normalized spectrum and the filter profile for the process of convolution. Convolution is necessary to obtain part of the spectrum permitted by the filter. This process requires pixel-to-pixel multiplication of the de-normalized flux spectrum and the filter profile, which is formulated as \(F_{\rm V}=\int_{0}^{\infty}S_\lambda(V)F_\lambda d\lambda\) in Equation~(\ref{equ:5}), where $F_\lambda = \pi B_\lambda(T_{\rm eff})$. One of the numerical techniques could be used to find the visual signal $F_{\rm V}$ after the convolution.

Performing the numerical integral \(F=\int_{0}^{\infty}F_\lambda d\lambda = \sigma T_{\rm eff}^4\) over the de-normalized flux spectrum is always problematic because of its limited range in wavelengths. Fortunately, this integral could be avoided by the definition of effective temperature that assures the area under the total de-normalized flux spectrum is equal to the area under the total real flux spectrum of the star. At last, the visual to bolometric luminosity ratio \((L_{\rm V}/L)\) is obtained as

\begin{equation}
    \frac{L_{\rm V}}{L}=\frac{\int_{0}^{\infty}S_\lambda(V)F_\lambda d\lambda}{\int_{0}^{\infty}F_\lambda d\lambda}=\frac{F_{\rm V}}{\sigma T_{\rm eff}^4}
    \label{equ:A9}
\end{equation}

\noindent after dividing the visual signal (\(F_{\rm V}\)) by the bolometric signal (\(\sigma T_{\rm eff}^4\)) even if the radius \((R)\) of star is unknown.

There could be three types of uncertainties that contribute to the uncertainty of \(L_{\rm V}/L\). Assuming that the integration and the truncation errors are negligible along with the uncertainties of visual ($L_{\rm V}$) and bolometric ($L$) signals, and the errors in the visual and bolometric signals are about the same and both are characterized by the \textit{S/N} of each spectrum, the relative error of \(L_{\rm V}/L\) would be estimated as

\begin{equation}
    \frac{\Delta \left(L_{\rm V}/L\right)}{\left(L_{\rm V}/L\right)}=\sqrt{\left(\frac{\Delta L_{\rm V}}{L_{\rm V}}\right)^2+\left(\frac{\Delta L}{L}\right)^2}=\frac{\sqrt{2}}{S/N}
    \label{equ:13}
\end{equation}

\noindent Consequently, Equation (\ref{equ:5}) implies that the error propagation up to zero-point constant $C_{\rm 2}$ of the  {\it BC} scale could be completed for a star as

\begin{equation}
    \Delta C_{\rm 2}=\sqrt{\left(\Delta BC_{\rm V}\right)^2+\left(1.0857\times\frac{\sqrt{2}}{S/N}\right)^2}
    \label{equ:14}
\end{equation}
where the numerical value ``$2.5\times \log_{10}{e}=1.0857$" is for converting the relative error of \(L_{\rm V}/L\) into magnitude scale because classically computed $BC_{\rm V}$ from $M_{\rm Bol} - M_{\rm V}$ and its uncertainty ($\Delta BC_{\rm V}$) is already expressed in magnitudes.

Equations (\ref{equ:13}) and (\ref{equ:14}) do not contain the term $T_{\rm eff}$. Is the error contribution of $T_{\rm eff}$ ignored? No, it is not. It is included in $L$ together with the effect of $R$. The effect of $R$ is canceled in $L_{\rm V} / L$, so it is possible to write it as $F_{\rm V} / F$, the visual to bolometric flux ratio on the surface of the star. Expressing this quantity in terms of fractions, that is, division of $d(L_{\rm V} / L)$ by ($L_{\rm V} / L$), further reduces the effect of $T_{\rm eff}$; thus, $BC_{\rm V}$ is not the same at all temperatures while Eqaution~(\ref{equ:A9}) is free from $R$.

Steps one and two must be repeated with different stars for numerous independent estimates of \(C_{\rm 2}\), from which a weighted or arithmetical mean of $\langle C_{\rm 2}\rangle$ and its associated error may be computed according to Equation (\ref{equ:5}). These could then be used to estimate \(C_{\rm V}\) and \(c_{\rm V}\) for the absolute and apparent visual magnitudes, such as:

\begin{equation}
    C_{\rm V} = (71.197425... - \langle C_{\rm 2}\rangle) \pm \,\,\text{S.E., if \(L_{\rm V}\) is in SI units}
    \label{equ:15}
\end{equation}

\begin{equation}
    c_{\rm V} = (-18.997351... - \langle C_{\rm 2}\rangle) \pm \,\,\text{S.E., if \(f_{\rm V}\) is in SI units}
    \label{equ:16}
\end{equation}

\noindent Having \(C_{\rm V}\) determined, one can use Equation~(\ref{equ:4}) to calculate \(L_{\rm V}\) for a star from its absolute visual magnitude, or having \(c_{\rm V}\) determined, one can use: 

\begin{equation}
    V = 2.5\log{f_{\rm V}}+c_{\rm V}
    \label{equ:17}
\end{equation}

\noindent to convert the visual flux (Wm$^{-2}$) arriving at Earth into apparent visual magnitude or vice versa. 

\section{A method of obtaining spectroscopic \texorpdfstring{$BC_{\rm V}$}{BCV} and \textit{V} Band Interstellar Extinction}

Once sufficiently accurate zero-point constants (\(C_{\rm V}\) and \(c_{\rm V}\)) of absolute and apparent visual magnitudes and/or the zero-point constant of the \(BC_{\rm V}\) scale \(C_{\rm 2}\) were determined by the method described above, obtaining a spectroscopic \(BC_{\rm V}\) and then an interstellar extinction (\(A_{\rm V}\)) in the visual band for a star from its high-resolution spectrum becomes possible. Only the effective temperature ($T_{\rm eff}$) of the star is required as a pre-determined quantity.

The equation to be used is 

\begin{equation}
    BC_{\rm V}=2.5\log{\frac{L_{\rm V}}{L}}+C_{\rm 2}=2.5\log{\frac{\int_{0}^{\infty}S_\lambda(V) F_\lambda d\lambda}{\sigma T_{\rm eff}^4}}+C_{\rm 2}
    \label{equ:18}
\end{equation}

\noindent where the visual to bolometric luminosity ratio (${L_{\rm V}/L}$) could be computed as a division of the visual signal, which is a quantity obtained by a numerical integration over the de-normalized flux spectrum convoluted by the profile function $S_\lambda(V)$, by the bolometric signal, which is the flux of a blackbody with a temperature $T_{\rm eff}$.

Unlike the uncertainty of the photometric \(BC_{\rm V}\) computed from $M_{\rm Bol}$ and $M_{\rm V}$ having numerous contributions such as from the effective temperature ($T_{\rm eff}$), radius ($R$), trigonometric parallax ($\varpi$), apparent magnitude ($V$) and interstellar extinction ($A_{\rm V}$), the uncertainty of the spectroscopic \(BC_{\rm V}\) has mainly two sources which are the visual and the bolometric signals characterized by the \textit{S/N} ratio of the observed spectrum. Thus, the error on the spectroscopic \(BC_{\rm V}\) is expected to be much smaller than the error of the photometric \(BC_{\rm V}\) of a star. Equation (\ref{equ:13}) gives it in the form of a fraction (per cent error) if the error contribution of the zero point constant $C_2$ is negligible. Multiplying it by the number ``$2.5\times \log_{10}{e}$" changes it into magnitude units.

After estimating the spectroscopic \(BC_{\rm V}\) of the star and its uncertainty as described above, the following equation could be used to calculate its interstellar extinction, \(A_{\rm V}\).

\begin{equation}
    A_{\rm V}=V+5\log{\varpi}+5-\left(M_{\rm Bol}-BC_{\rm V}\right)
\label{equ:19}
\end{equation}

\noindent where \(M_{\rm Bol}\) must be calculated according to Equation~(\ref{equ:3}) from \(L\) of star in SI units. The term \(\left(M_{\rm Bol}-BC_{\rm V}\right)\) is equivalent to the absolute visual magnitude \((M_{\rm V})\) according to Equation~(\ref{equ:5}). It is clear in this equation that if the observed absolute visual magnitude, which is \(\left(V+5\log{\varpi}+5\right)\), is equal to the computed absolute visual magnitude, which is \(\left(M_{\rm Bol}-BC_{\rm V}\right)\), there is no extinction. But, if the observed absolute visual magnitude is fainter (larger value) than the computed absolute visual magnitude, one will compute a positive value for \(A_{\rm V}\), which means that there must be an interstellar extinction for this star. A negative value for \(A_{\rm V}\) is physically impossible, meaning that there is no interstellar extinction.

\end{document}